\documentclass[journal,twocolumn]{IEEEtran}

\usepackage[svgnames]{xcolor}

\usepackage{braket}
\usepackage{multirow}
\usepackage{float}
\usepackage{soul}
\usepackage{blkarray}
\usepackage{multicol}
\usepackage{hyperref}

\DeclareRobustCommand\openone{\leavevmode\hbox{\small1\normalsize\kern-.33em1}}

\usepackage{cite}

   \usepackage{graphicx}

\usepackage{amsmath}
\usepackage{array}

\usepackage[caption=false,font=footnotesize]{subfig}
\DeclareRobustCommand*{\IEEEauthorrefmark}[1]{%
	\raisebox{0pt}[0pt][0pt]{\textsuperscript{\footnotesize #1}}%
}

\usepackage{tikz}
\usetikzlibrary{decorations.markings}
\usetikzlibrary{shapes.geometric}
\usetikzlibrary{fit}

\newcommand{\overarrowscale}{0.5} 
\newcommand{\boxscalefactor}{1.2} 

\pgfdeclarelayer{edgelayer}
\pgfdeclarelayer{nodelayer}
\pgfsetlayers{edgelayer,nodelayer,main}

\tikzstyle{none}=[inner sep=0pt]

\tikzstyle{rn}=[circle,fill=Red,draw=Black,line width=0.8 pt]
\tikzstyle{gn}=[circle,fill=Lime,draw=Black,line width=0.8 pt]
\tikzstyle{yn}=[circle,fill=Yellow,draw=Black,line width=0.8 pt]
\tikzstyle{data_qubit}=[regular polygon,regular polygon sides=3,shape border rotate=0,fill=RoyalBlue,draw=Black]
\tikzstyle{parity_check_qubit}=[shape=star,star points=5, star point ratio=2.25,fill=Red,draw=Black]
\tikzstyle{classical_parity_check}=[rectangle,fill=White,draw=Black,scale=2]
\tikzstyle{classical_data_bit}=[circle,fill=Gold,draw=Black,scale=0.8]
\tikzstyle{classical_data_phase}=[circle,fill=DeepSkyBlue,draw=Black,scale=0.8]
\tikzstyle{classical_data}=[circle,fill=White,draw=Black,scale=0.8]
\tikzstyle{arrow_end}=[circle,fill=none,draw=none,scale=.1]
\tikzstyle{X}=[circle,fill=White,draw=none,label={[xshift=0 cm, yshift=-0.3 cm,Grey]X},scale=.1]
\tikzstyle{Z}=[circle,fill=White,draw=none,label={[xshift=0 cm, yshift=-0.3 cm,Grey]Z},scale=.1]
\tikzstyle{a}=[circle,fill=none,draw=none,scale=.1,label={[xshift=0 cm, yshift=-0.5 cm]a)}]
\tikzstyle{b}=[circle,fill=none,draw=none,scale=.1,label={[xshift=0 cm, yshift=-0.5 cm]b)}]
\tikzstyle{c}=[circle,fill=none,draw=none,scale=.1,label={[xshift=0 cm, yshift=-0.5 cm]c)}]
\tikzstyle{eq}=[circle,fill=none,draw=none,scale=2,label={[xshift=0 cm, yshift=-0.5 cm]=}]
\tikzstyle{cir}=[circle,fill=none,draw=Blue,dashed,scale=4,line width=2 pt]
\definecolor{tempcolor}{rgb}{.9,.7,0}
\tikzstyle{cirX}=[circle,fill=none,draw=tempcolor,scale=4,line width=2 pt]
\tikzstyle{cirZ}=[circle,fill=none,draw=DeepSkyBlue,scale=4,line width=2 pt]

\tikzstyle{simple}=[-,draw=Black,line width=1.000]
\tikzstyle{arrow}=[-,draw=Black,postaction={decorate},decoration={markings,mark=at position .5 with {\arrow{>}}},line width=2.000]
\tikzstyle{tick}=[-,draw=Black,postaction={decorate},decoration={markings,mark=at position .5 with {\draw (0,-0.1) -- (0,0.1);}},line width=2.000]
\tikzstyle{conj_prop}=[-,draw=Black,line width=1.00]
\tikzstyle{CNOT}=[-,draw=Red,line width=1.00]
\tikzstyle{propagation}=[-,draw=Magenta,postaction={decorate},decoration={markings,mark=at position .5 with {\arrow{>}}},line width=1.00]
\tikzstyle{big_arrow}=[-latex,draw=Black,line width=3.000]
\tikzstyle{thin_arrow}=[->,draw=Grey,dash pattern=on \pgflinewidth off 2pt,line width=1]
\tikzstyle{undetected}=[dashed,postaction={decorate},decoration={markings,mark=at position .5 with {\arrow{>}}},draw=Grey,line width=1.5]
\tikzstyle{arrow_grey}=[->,draw=Grey,dash pattern=on \pgflinewidth off 2pt,line width=1]
\tikzstyle{added_edge}=[-,draw=Blue,line width=1.5]
\tikzstyle{cancelled_edge}=[-,draw=none,dashed,line width=1.5]
\tikzstyle{indirect_prop}=[-,draw=Blue,line width=1.5]
\definecolor{tempcolor}{rgb}{.9,.7,0}
\tikzstyle{added_S1}=[-,draw=tempcolor,line width=1.5]
\tikzstyle{added_S2}=[-,draw=DeepSkyBlue,line width=1.5]

\usepackage[textsize=footnotesize,backgroundcolor=red!25]{todonotes}

\def\beq{\begin{eqnarray}}
\def\eeq{\end{eqnarray}}

\newcommand{\synd}[2]{{\textbf{\color{red}#1}}{\textbf{\color{cyan}#2}}}

\begin{document}
%
\title{Quantum Codes from Classical\\Graphical Models}

\author{
	\IEEEauthorblockN{Joschka~Roffe\IEEEauthorrefmark{1,4}, Stefan~Zohren\IEEEauthorrefmark{2}, Dominic~Horsman\IEEEauthorrefmark{3,1}, Nicholas~Chancellor\IEEEauthorrefmark{1}}
\\ \IEEEauthorblockA{\IEEEauthorrefmark{1}\textit{Joint Quantum Centre (JQC) Durham-Newcastle, Department of Physics, Durham
	University, United Kingdom}}
\\ \IEEEauthorblockA{\IEEEauthorrefmark{2}\textit{Oxford-Man Institute, Department of Engineering Science, Oxford University, United Kingdom}}
\\ \IEEEauthorblockA{\IEEEauthorrefmark{3}\textit{Laboratoire d'Informatique de Grenoble, Universit\'e Grenoble Alpes, France}}
\\ \IEEEauthorblockA{\IEEEauthorrefmark{4}\textit{Department of Physics \& Astronomy, University of Sheffield, United Kingdom}}
\thanks{CONTACT Joschka Roffe: joschka@roffe.co.uk}
\thanks{\copyright \ 2019 IEEE.  Personal use of this material is permitted.  Permission from IEEE must be obtained for all other uses, in any current or future media, including reprinting/republishing this material for advertising or promotional purposes, creating new collective works, for resale or redistribution to servers or lists, or reuse of any copyrighted component of this work in other works.}
}

\maketitle

\begin{abstract}
We introduce a new graphical framework for designing quantum error correction codes based on classical principles. A key feature of this graphical language, over previous approaches, is that it is closely related to that of factor graphs or graphical models in classical information theory and machine learning. It enables us to formulate the description of the recently-introduced `coherent parity check' quantum error correction codes entirely within the language of classical information theory. This makes our construction accessible without requiring background in quantum error correction or even quantum mechanics in general. More importantly, this allows for a collaborative interplay where one can design new quantum error correction codes derived from classical codes. 
\end{abstract}


\IEEEpeerreviewmaketitle

\section{Introduction and background}

\IEEEPARstart{I}{nformation} is processed, communicated and stored using physical systems that are susceptible to error. As such, error detection and correction protocols are necessary to ensure reliable operation. The fundamental principle underpinning classical information theory and error correction \cite{Cover:2006,MacKay03a} is that data is redundantly encoded across an expanded space of bits. The resultant \emph{logical} data has additional degrees of freedom which can be exploited to actively detect and correct errors. The exact method by which information is redundantly encoded to create logical data is specified by a set of instructions know as a \emph{code}. In practice, most error correction schemes are based on an efficient class of protocols known as linear \emph{block} codes. For block codes, error correction proceeds by tracking the correlations between data bits by  using parity checks. The role of the additional redundancy bits in a block code is to store the parity information so that it can be decoded over time. Modern protocols such as low-density-parity check (LDPC) codes \cite{Gallanger63,MacKay1996,MacKay1999} and turbo codes \cite{Berrou1993,Berrou1996} perform at close to the Shannon rate, which is the maximum theoretical rate for information transfer along a noisy channel \cite{Shannon}.

In quantum error correction \cite{Shor1995,Stea1996b,Cald1995,Lida2013} bits are replaced with quantum bits (from now on referred to as \emph{qubits}). Qubits exhibit several features, discussed in more detail below, which complicate the process of creating quantum error correction codes. State of the art quantum error correction codes, such as the surface code \cite{fowler2012surface}, rely upon a special type of quantum measurement known as a stabilizer \cite{Gottesman:1998hu}. Stabilizers play a similar role to the previously mentioned parity checks, but are subject to various constraints that make it difficult to derive quantum codes in direct analogy to efficient classical codes.   

In recent work \cite{chancellor16}, the so-called coherent parity check (CPC) codes were introduced as a new framework to derive quantum error correction codes. The specific advantage of CPC codes is a \emph{fail-safe} code structure that guarantees that the quantum mechanical requirements of the code are satisfied. As a result, the CPC construction provides a useful framework for the conversion of classical block codes to quantum codes. In \cite{chancellor16}, CPC code design was demonstrated using the {\sc{zx}}-calculus as well as a corresponding generator matrix formulation.

In this work, we outline a complementary, special-purpose, formalism for CPC codes to allow them to be expressed in terms of classical factor graphs of the type commonly used in classical information theory and machine learning \cite{Kschischang2001}. Factor graphs provide a simple representation of correlations within multivariate probability distributions. In the context of coding theory, factor graphs reveal structure that enables decoding using efficient approximate inference algorithms \cite{MacKay1996,MacKay1999,MacKay03a}. Factor graph techniques have previously been adapted for the decoding of quantum codes in \cite{Leifer2008,Renes2017}, as well as being used as tools for the discovery and analysis of new quantum codes in \cite{Vontobel2008,Li2016}. In this paper, we develop factor graph methods for use with CPC codes. The strength of our mapping is that it provides a general tool for optimization of quantum codes which does not require the user to be versed in quantum theory. Such flexible tools are likely to be important in the near term, as gate model quantum devices are just entering the so-called `noisy intermediate-scale' stage of their development \cite{Preskill2018}. Given the highly constrained nature of these early quantum devices, bespoke error correction protocols will be required. As such, there is a need for optimization and design methods capable of incorporating a variety of hardware constraints. Our methods here are intended to fill precisely this niche, allowing codes to be optimised against complex metrics for quality and suitability for specific hardware.

In developing the factor graph mapping for CPC codes, we first introduce an intermediate graphical language called the operational representation. In addition to providing a visualisation of operations performed between qubits, the role of the operational representation is to abstract away the quantum mechanical properties of the code into a form that enables it to be mapped to a classical factor graph. Following this translation, any code design and optimisation proceeds using purely standard graphical models, without any further reference to quantum mechanics. We first demonstrate the utility of the classical factor graph representation for quantum code design by outlining the derivation of a simple quantum detection code. We then describe how factor graphs can be used to construct a quantum error correction code based on classical Hamming codes.

\subsection{Graphical models in classical information theory} \label{sec:classical}

\begin{figure}
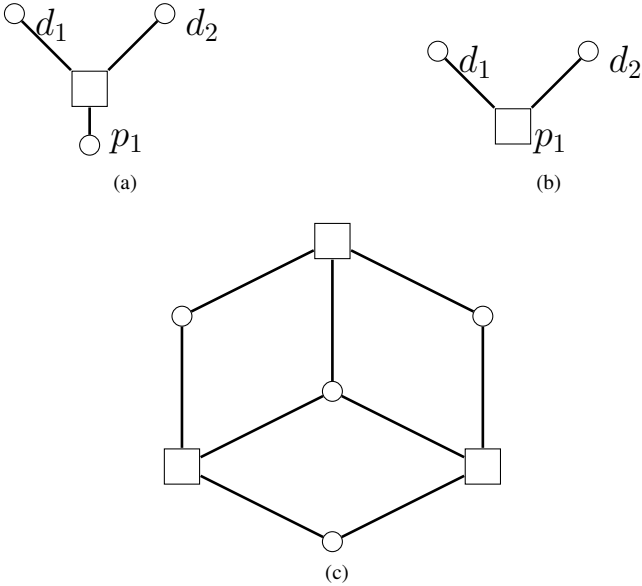
 
	\centering
	\subfloat[]{%
			\input{figures/1a.tikz}  }
	\hfill
	\subfloat[]{%
			\input{figures/1b.tikz}  }
	\hfill
	\subfloat[]{%
		\input{figures/1c.tikz}}
	\caption{Examples of classical factor graphs. (a) The $[3,2,2]$ detection code in standard factor graph notation including `dangling' edge to indicate measurement of the parity.  (b) The same but with the additional node suppressed for visual clarity, this is the style which we use for the remainder of the manuscript, with a single additional node implicit on each parity check node. (c) The $[7,4,3]$ Hamming code in our notation. Note that in the context of coding these graphs (without the node suppressed) are sometimes referred to as Tanner graphs.}
	\label{fig:cfg1}
\end{figure}

We begin by outlining standard conventions in the graphical representation of classical error correction codes (for an overview see \cite{MacKay03a,Koller:2009}). This provides a point of reference with which to compare the graphical language for quantum error correction presented in this paper. In a classical block code, parity bits are introduced to measure and track the parity of the data bits. A factor graph is a tool designed to provide a visualisation of the relationship between data and parity bits in a given error correction code.

As a simple first example of factor graph notation, we consider a single (classical) parity-check cycle on a two-bit data register $\mathbf{r}=\left(D_1,D_2\right)$, where $D_1$ and $D_2$ are classical data bits with values $0$ or $1$. In the encode stage of the parity-check cycle, an extra bit is introduced to measure and store the \emph{parity check} of the data bits, which is calculated as the binary sum $p=\left(D_1+D_2\right)\!\mathrm{mod} \ 2$. The resultant three-bit string $\mathbf{c}=\left( D_1,D_2,p \right)$ is called the codeword. A factor graph representation of this encoding is shown in Figure \ref{fig:cfg1}a, where the circles represent the data bits and the square the parity bit. The edges in the factor graph indicate which data bits are involved in a given parity check. For the graphs in this paper we choose to suppress the edge and node indicating the measurement of each parity check as depicted in Figure \ref{fig:cfg1}b. Since every check will have a single such node, doing so does not introduce any ambiguity in our notation. Mathematically, we can also express this relation through the \emph{generator matrix} of the code $\mathbf{G}$, which relates the data register and the code word via $\mathbf{c} = \mathbf{G}^T \mathbf{r}$ ($\mathrm{mod} \ 2$). In the above example the generator matrix reads
\begin{equation} \label{eq:generator}
\mathbf{G} = \left[\begin{array}{ccc}1 & 0 & 1\\0 & 1 & 1 \end{array}\right]
\end{equation}
giving the code word equation
\begin{equation} \label{eq:generator}
\left[\begin{array}{c} D_1 \\ D_2 \\ p \end{array}\right]  = \left[\begin{array}{cc}1 & 0 \\0 & 1 \\1 & 1\end{array}\right]  \left[\begin{array}{c} D_1 \\ D_2 \end{array}\right] =\left[\begin{array}{c} D_1 \\ D_2 \\ D_1 + D_2 \end{array}\right]
\end{equation}
Whilst in some cases it is useful to work with the generator matrix, we note that the same information is contained in the factor graph, on which we focus here.

A single error on any of the three bits in the protocol depicted in Figure \ref{fig:cfg1}a,b can be detected by comparing the values of parity checks at successive times $p(t_0)$ and $p(t_1)$. A bit-flip error occurring between these checks will cause the value of the parity to change such that $p(t_0)\neq p(t_1)$. Under the standard labelling convention, this parity-check cycle is an $[n=3,k=2,d=2]$ code, where $n$ is the total number of bits and $k$ is the number of data bits. The code distance $d$ is the Hamming distance between code words, and so is the minimum weight of an error operator that will not be picked up as an error. For the code depicted in Figure \ref{fig:cfg1}a, the distance is $d=2$, as an error of weight two will cause the parity bit to flip back to its original value.

The $[3,2,2]$ code can detect the presence of a single bit-flip, but does not provide enough information to pinpoint which of the bits the error occurred on. It is therefore a detection code. Full error correction codes, with the ability to both detect and localise errors require multiple overlapping parity checks. An example is the $[7,4,3]$ Hamming code, the factor graph for which is shown in Figure \ref{fig:cfg1}c. For correction codes, the task of decoding involves deducing the most likely error from a set of parity-check measurements. This information about the most likely error allows the data to be corrected to its original state with high probability. For the case of small codes, such as the Hamming code, decode tables can be constructed by exhaustively testing the code with all possible error chains. 

In general, to decode larger codes one views the decoding problem as a maximum posterior inference problem. The use of factor graphs naturally leads to techniques from graphical models \cite{Koller:2009} to efficiently find the maximum posterior estimator. In many large real world codes, doing exact maximum posterior inference is computationally hard, but efficient approximate inference algorithms are known, such as belief propagation \cite{Koller:2009}. Therefore, the use of graphical model representations of error correction codes not only provides an intuitive visualisation of the generator matrix, but also affords access to a number of powerful approximate inference algorithms designed for graphical models.

\subsection{From classical to quantum error correction}

\begin{figure}
	\begin{center}
		\includegraphics[height=5cm]{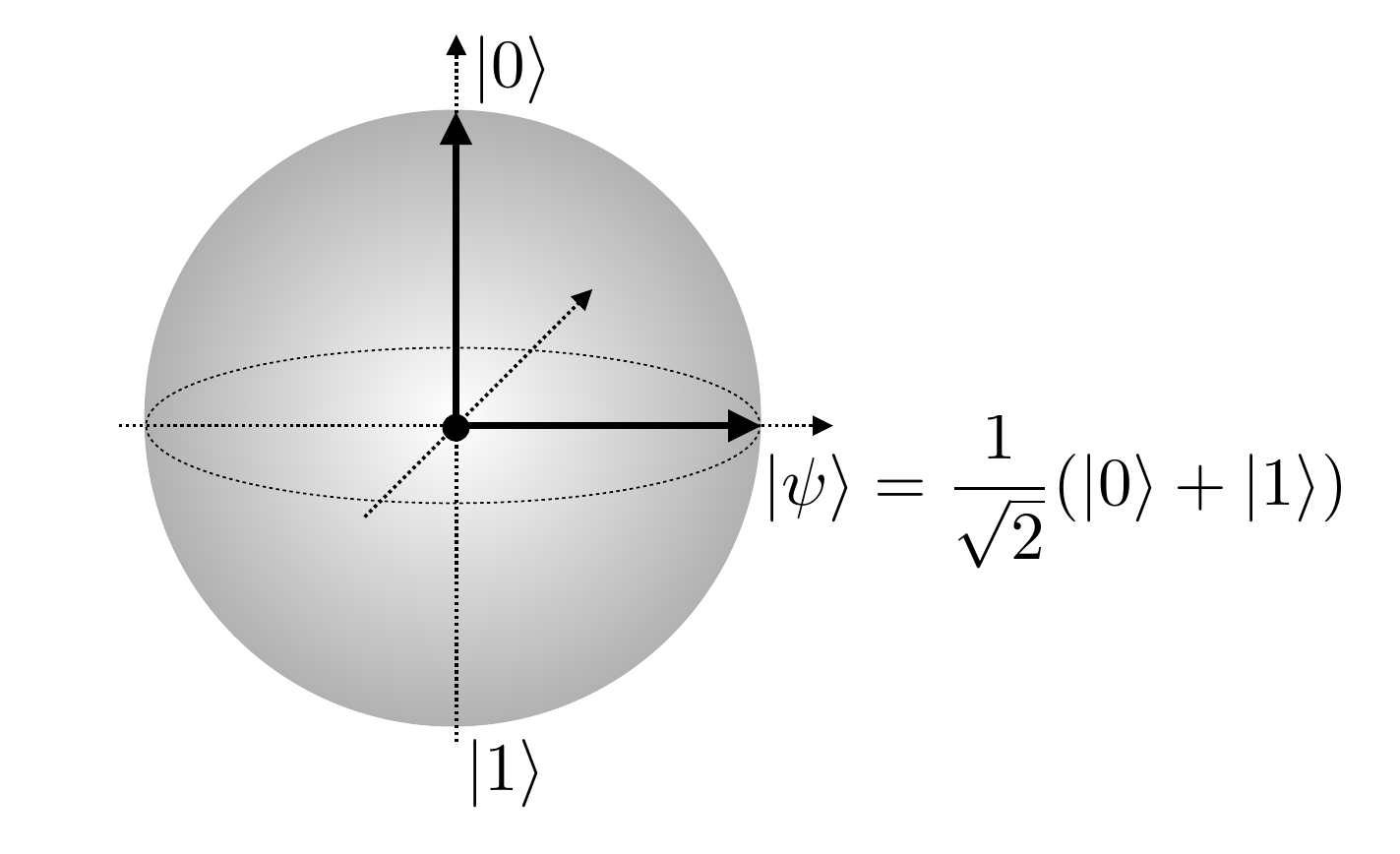}
		\caption{Pure quantum states can be represented as a point on the surface of the so-called Bloch sphere.}
		\label{fig:bloch}
	\end{center}
\end{figure}

Before we describe our graphical language for quantum error correction, we give a brief overview of the challenges quantum mechanics imposes on quantum code design (for reviews see \cite{Nielsen,Rieffel}). This discussion is included to put our work into context, as our graphical language is self-consistent and can in principle be applied without any knowledge of quantum mechanics.

In classical information theory and computer science, information is commonly represented through bits which take values $0$ and $1$. In quantum mechanics, the formalism gives rise to what is known as a \emph{quantum bit}, or \emph{qubit} for short. The mathematical representation of a qubit can best be understood as a point on a Bloch sphere (see Figure \ref{fig:bloch}), which represents a \emph{wavefunction} denoted as $\ket{\psi}$ in Dirac notation \cite{DiracPrinciples}. Analogous to classical bits, a qubit can be in states $\ket{\psi}=\ket{0}$ and $\ket{\psi}=\ket{1}$, as depicted in Figure \ref{fig:bloch}. However, they can also be in a superposition such as $\ket{\psi} = \frac{1}{\sqrt{2}}(\ket{0}+\ket{1})$. The general form of the qubit wavefunction is given by $\ket{\psi}=\alpha\ket{0}+\beta\ket{1}$ where $\alpha$ and $\beta$ are complex numbers satisfying the condition $|\alpha|^2+|\beta|^2=1$. One important property of quantum mechanics is known as the \emph{collapse of the wave function}. A measurement of the general state 
\beq \label{eq:generalstate}
\ket{\psi}=\alpha\ket{0}+\beta\ket{1}, 
\eeq
for example, `collapses' the qubit to $\ket{\psi}=\ket{0}$ or $\ket{\psi}=\ket{1}$ with probabilities $|\alpha|^2$ and $|\beta|^2$ respectively. The important consequence is that measurement is not typically passive in quantum mechanics: measuring a system in general changes its state. The above is enough to understand some of the challenges of quantum error correction.

First, as a qubit is no longer a binary number but the mathematical equivalent of a point on a three-dimensional sphere, it becomes clear that errors can occur in different forms, corresponding to flipping the state around a different axis. This gives rise to two different types of errors, the so-called \emph{bit and phase errors}. The standard quantum notation for a bit-flip operator is the symbol $X$, which has the following effect on the general qubit state \eqref{eq:generalstate}
\beq
X\ket{\psi}=\alpha X\ket{0}+\beta X\ket{1} =\alpha\ket{1}+\beta\ket{0}. 
\eeq
Similarly, phase-flip operators are represented by the symbol $Z$ and transform the general qubit state \eqref{eq:generalstate} as follows
\beq
Z\ket{\psi}=\alpha Z\ket{0}+\beta Z\ket{1}= \alpha\ket{0}-\beta\ket{1}.
\eeq 
In this paper, we also refer to bit-flip and phase-flip errors as \emph{X-errors} and \emph{Z-errors} respectively. Mathematically, the operators $\{X,Z\}$ can be represented as Pauli matrices. A quantum error correction code must have the ability to detect and correct both types of errors. Technically, there is also the possibility of a $Y$-error which can be understood as a simultaneous $X$- and $Z$-error.  For our outline of graphical models for quantum error correction, it initially suffices to focus on $X$- and $Z$-errors only, with discussion of $Y$-errors left to the end. Note that because of the projective nature of quantum measurement, the ability to determine the location of $X$, $Z$, and $Y$ errors is sufficient to correct against arbitrary quantum errors.

A second issue is that arbitrary quantum data cannot be copied, or \emph{cloned}. In general, then, simple quantum repetition codes cannot be established \cite{Park1970,Wootters1982}. 

A third challenge in the design of quantum error correction codes has to do with the aforementioned `collapse of the wave function'.
Parity checking depends crucially on measurement, which is considered non-disturbing classically. In contrast, the parity checking sequences in a quantum code must be carefully chosen so not to collapse the encoded information. Such non-disturbing measurements are referred to as \emph{stabilizers} \cite{Gottesman:1998hu}.

A final complication to quantum error correction compared to classical error correction is that quantum parity checks are performed via `unitary' operations on the combined system of data plus parity qubits. A consequence of unitarity is that operations are in general bi-directional: both qubits involved in an operation change their state (this is connected to the fact that quantum data cannot be cloned). This means that parity checks themselves can propagate faults to the data register, resulting in an additional pathway for errors that needs to be accounted for.

\subsection{Coherent parity check codes}

\begin{figure}
	\centering
	\scalebox{.8}{\input{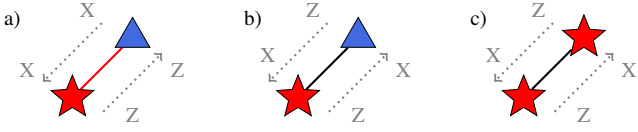}}
	\caption{The operational representation: The triangle nodes represent data qubits and the stars parity qubits. The nodes are connected by three types of edges. (a) Bit-check edges (drawn in red) copy $X$-errors from data qubits to parity qubits. $Z$-errors are back propagated in the reverse direction. (b) Phase-check edges (drawn in black) between data and parity qubits. This edge propagates a $Z$-error as a $X$-error between the qubits. The phase-check edge is symmetric and has the same error propagation behaviour in both directions. (c) Cross-check edges connect parity check qubits to other parity check qubits. The error propagation behaviour for cross-check edges is identical to that of phase-check edges.}
	\label{fig:fg_edges}
\end{figure}

The challenges of quantum error correction described in the previous section have complicated the development of good quantum codes with high rates. For example, the surface code \cite{nqit_nickerson,Barends2014,Takita2017,fowler2012surface} -- currently the favoured experimental quantum error correction protocol -- requires a minimum of 13 qubits \footnote{The minimum number of qubits is 10 if the requirement for nearest-neighbour-only interactions is dropped.} to encode a single logical qubit \cite{lattice-surgery}. The coherent parity check (CPC) framework for quantum error correction has recently been introduced in \cite{chancellor16} and further developed in \cite{roffe17}, with the aim of providing a toolset for the construction of efficient quantum codes as well as easily modify existing high-quality codes. Central to this new framework is a fail-safe code structure that ensures all CPC codes are stabilizer codes. The advantage to this is that there no longer restrictions on the form of parity checks performed on the quantum data; the fact that CPC codes are stabilizer codes guarantees that the chosen parity check will not collapse the encoded information. Consequently, the CPC framework allows for code construction methods that do not rely upon a detailed understanding of quantum mechanics. In \cite{roffe17}, it is shown how CPC codes can be discovered via exhaustive machine search. Another promising application of the CPC framework is as a tool for the conversion of classical codes to quantum codes \cite{chancellor16}. Here, we expand upon this work by demonstrating that CPC codes can be described using graphical methods inspired by the classical factor graph notation.

As is the case in classical parity check codes such as the Hamming code, the qubits in a CPC code are separated into data qubits and parity check qubits. Error detection in a CPC code involves performing a round of phase-parity checks, followed by a round of bit-parity checks. At the end of the code a final round of \emph{cross-checks} is performed between the parity check qubits to mitigate the effect of undetected errors on the parity check qubits. The only requirement for this construction is that the two classical codes encode the same number of bits.
A more detailed presentation of the CPC code structure can be found in \cite{chancellor16,roffe17}, although we give an overview of the structure of the codes in circuit notation in appendix \ref{app:qcm}.  Note that the order in which the different types of parity check are performed in a CPC encoder is important. In the following sections we assume a canonical CPC ordering involving cross-checks, bit-checks and then phase-checks. This ensures that there is a restricted set of error propagation pathways that can be accounted for in a systematic way.

\subsection{CPC and CSS codes}

We now discuss the relationship between our CPC code design methods and the existing Calderbank, Shor, and Steane (CSS) \cite{Cald1995, Stea1996b} methods for stabilizer code construction. The CSS construction provides a method by which a quantum code can be formed by combining a pair of dual classical codes. However,  as the number of classical codes satisfying the duality requirement is limited, the CSS construction does not provide a comprehensive method for the translation of arbitrary classical codes to quantum codes. In contrast, the CPC code design method guarantees a stabilizer code from any pair of classical codes that encode the same number of logical bits. It is not, in fact, constructive -- unlike our CPC code design method. This guarantees a stabilizer code from any pair of classical codes that encode the same number of logical bits (while noting that the resultant stabilizer code has a reduced code distance compared to the original classical code). As a result, the CPC framework can be seen as a better tool for making contact between classical and quantum coding theory.

In \cite{chancellor16} it is proved that any CSS code can be expressed as a CPC code with a three-part encoder. Furthermore, a CPC structural template is presented in \cite{Roffe_thesis} that allows a distance-three CSS code to be derived from the starting point of (almost) any distance-three classical code. As a result, the CPC methods for distance-three code construction improves over the original CSS approach. It should also be noted that the CPC framework is not limited to CSS codes. For example, the $[[10,4,3]]$ code derived in Section \ref{sec:10_4_3} falls outside the CSS family of codes.

Although all the CPC codes in this paper are derived from the starting point of classical codes, there are other methods for CPC code design. In \cite{roffe17}, for example, it is shown that new CPC codes can be discovered via machine search techniques. Another approach to CPC code design is to start with an existing quantum code, and use the graphical methods outlined in this paper to modify it. In contrast, the CSS construction does not include the flexibility to modify existing codes in this way.

When a CPC code is constructed from two classical codes, the resultant code will not be the same as a CSS code constructed from those two codes (assuming they meet the necessary conditions). This can be most easily seen by a counting argument on the number of logical qubits. The number of logical qubits in a CPC code is equal to the number of logical bits in each classical code. In contrast, for a CSS code, the number of logical qubits is equal to $|k_1-k_2|$, where $k_1$ and $k_2$ are the number of logical qubits in each of the input classical codes \cite{Cald1995,Stea1996b}.
 
 A further, notable, advantage of the CPC construction is that it comes equipped with high-level graphical languages and representations -- including those given in this paper. The lack of high-level tools has been a significant bottleneck in the development of deployable quantum codes.
 The CPC construction includes CSS codes, but it a much broader and more powerful construction for general stabilizer codes.

 Other methods for constructing quantum codes from arbitrary pairs of classical codes include entanglement assisted codes \cite{Brun2006} and hypergraph product codes \cite{Tillich2014}. The strength of the CPC framework over the entanglement assisted construction is that it is not necessary to prepare additional noiseless entanglement bits as part of the CPC protocol. The advantage of CPC codes compared to hypergraph product codes is that it is possible to create smaller quantum codes relative to the size of the original classical base-code; the length of a CPC code is linearly proportional to the length of its classical base-code, compared to the quadratic increase in block-length that results from the hypergraph product construction. The disadvantage of the CPC construction, compared to the aforementioned methods, is that there is no guarantee on the minimum distance of the quantum code when the classical base-code has distance $d>3$.

\section{A graphical language for quantum parity check operations}

\subsection{Operational representation of quantum parity check codes}

We now introduce an intermediary graphical representation we call the operational representation of CPC codes. This notation is designed to enable easy visualisation of the propagation of the different types of quantum errors between qubits, and  will serve as a stepping stone to our eventual presentation of CPC codes using classical factor graphs in Section \ref{sec:qfg_to_cfg}.

Graphs of the operational representation have two types of nodes: 
\begin{enumerate}
	\item triangles (\scalebox{.5}{\begin{tikzpicture}
	\begin{pgfonlayer}{nodelayer}
		\node [style={data_qubit}] (0) at (0, -0) {};
	\end{pgfonlayer}
\end{tikzpicture}}) representing data qubits and
	\item stars (\scalebox{.5}{\begin{tikzpicture}
	\begin{pgfonlayer}{nodelayer}
		\node [style={parity_check_qubit}] (0) at (0, -0) {};
	\end{pgfonlayer}
\end{tikzpicture}}) representing parity qubits.
\end{enumerate}
The nodes are connected via edges that denote the different error propagation pathways between the qubits. The three types of edges between qubit nodes are shown in Figure \ref{fig:fg_edges}. The first type of edge in the operational representation, shown in red in Figure \ref{fig:fg_edges}a, is called a \emph{bit-check} edge and propagates quantum information in two directions: bit-errors are propagated from the data qubit to the parity qubit and phase-errors in the opposite direction. In a quantum computer, bit-check operations are implemented via the application of a controlled-not ({\sc cnot}) gate. Further details about the operation of {\sc cnot} gates in the context of CPC codes can be found in \cite{chancellor16,roffe17}.  Note that throughout this paper we assume that gates function ideally.

The black edge in Figure \ref{fig:fg_edges}b connects a data qubit to a parity qubit, and is referred to as a \emph{phase-check} edge. The phase-check edge propagates $Z$-errors and converts them into $X$-errors, as shown in Figure \ref{fig:fg_edges}b. This conversion between error types is important, as it allows a $Z$-error to be detected as a $X$-error via a parity check measurement. These gates also cause unavoidable propagation of $Z$-errors on the parity check qubits as $X$-errors on the data qubits.  Phase-check operations of this type are realised via the implementation of a conjugate-propagator gate in a quantum computer. Specific details about this gate are outlined in \cite{roffe17}.

\begin{figure}
	\centering
	\scalebox{.8}{\input{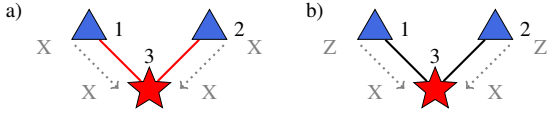}}
	\caption{The operational representation for quantum parity checking cycles on a register of two data qubits. The qubit nodes in these graphs are labelled 1-3. (a) The bit-flip code or $[[3,2,1]]$ code detects $X$-errors on any of the three qubits. $Z$-errors on the data qubits will, however, go undetected as the bit-check edges do not propagate $Z$-errors from data qubits to parity qubits. (b) The phase-flip code or $[[3,2,1]]$ code. Here $Z$-errors on the data qubits are propagated to the parity-qubit as $X$-errors, allowing them to be detected via a
		measurement. $Z$-errors on the parity qubit will propagate a correlated $XX$ error to the register that goes undetected.}
	\label{fig:321_codes}
\end{figure}

The third type of edge is a black edge connecting two parity qubits, as shown in Figure \ref{fig:fg_edges}c. This edge is called a \emph{cross-check}, and has the same error propagation behaviour as the phase-check operation. This gate is symmetric, so a $Z$-error on either qubit is propagated as a (detectable) $X$-error on the other.   Cross-check operations are useful for detecting the errors that are back-propagated to the register by the phase-check and bit-check edges.

Throughout this manuscript we will refer to individual edges (or two qubit operations) as bit, phase, or cross check edges (operations). This is not intended to imply that an individual one of these operations is performing a check, the parity checks themselves are actually performed by the qubit measurements which follow these interactions (and in general may involve both bit and phase information). We have chosen this terminology to highlight the role each of these interactions in terms of the kind of information they propagate.

Figure \ref{fig:321_codes} shows the operational representation for two $[[n=3,k=2,d=1]] = [[3,2,1]]$ quantum parity check codes; the first is designed to detect bit-flip errors and the second phase-flip errors. The error propagation rules summarised in Figure \ref{fig:fg_edges} can be used to determine the operation of each code. It can be verified that the bit-flip $[[3,2,1]]$ code will detect any single-qubit error from the set $\{X_1,X_2,X_3\}$. Here $X_i$ refers to the $X$-error of qubit $i$ as labelled in Figure \ref{fig:321_codes}. However, as we are dealing with a quantum information, the full single-qubit error set includes the phase-errors $\{Z_1,Z_2,Z_3\}$. Figure \ref{fig:321_codes}a shows that $Z$ errors on the data qubits in the bit-flip $[[3,2,1]]$ code are not propagated to the register. As a result a single-qubit error can go undetected, meaning the code has distance $d=1$.

Figure \ref{fig:321_codes}b shows the operational representation for the phase-flip $[[3,2,1]]$ code. Single-qubit $Z$-errors that occur on the data qubits are propagated to the parity qubit as an $X$-error, which is then detected by a 
measurement. A $Z$-error on the parity qubit itself, however, will go undetected as the parity qubit measurement is unchanged by phase-flip errors. This is a problem, as the black edges can propagate errors from the parity qubit back to the register, meaning certain errors can go undetected. When a $Z$-error occurs on the parity-qubit of the phase-flip $[[3,2,1]]$ code, a correlated $X_1X_2$ error is propagated to the register qubits. Back propagation of errors to the register in this way severely complicates the construction of quantum codes. In the next section, we show how the CPC codes can be designed to account for such errors.

\subsection{The [[4,2,2]] CPC detection code} \label{sec:422_code}

\begin{figure}
	\centering
{\centering	\scalebox{0.7}{\input{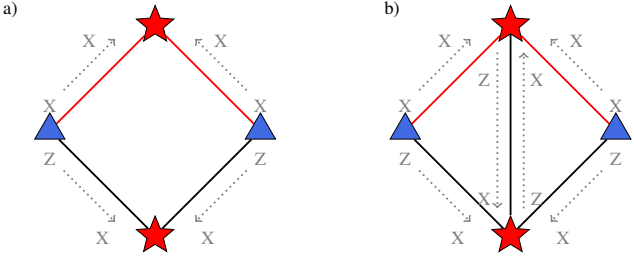}} }
	\caption{	Construction of a $[[4,2,2]]$ CPC detection code. (a) The operational representation for the code formed by combining a $[[3,2,1]]$ bit-flip code with a $[[3,2,1]]$ phase-flip code. A $Z$-error on the bit parity check qubit (labelled) will propagate errors to the register without flagging a check. As this error goes undetected, this is a $[[4,2,1]]$ code. (b) The $[[4,2,2]]$ code is constructed by adding a cross-check edge between the parity check qubits. The additional cross-check ensures all single-qubit errors are detected and fixes the code distance to $d=2$.}
	\label{fig:422_code}
\end{figure}

We now introduce the simplest CPC code, the $[[4,2,2]]$ detection code. This code is not only uniquely the smallest error detecting stabilizer code with two qubits \cite{Vaidman1996,Grassl1997}, it is also the smallest error detecting CPC code. The $[[4,2,2]]$ code is formed by combining the bit-flip $[[3,2,1]]$ code and the phase-flip $[[3,2,1]]$ code. Additional cross-check edges are then added between the parity qubits to ensure that any errors that are back-propagated to the register can be detected. This fixes the code distance to $d=2$ ensuring the $[[4,2,2]]$ code can detect both $X$- and $Z$-errors on any of the four qubits.

Figure \ref{fig:422_code}a shows the operational representation of a code constructed by combining the bit- and phase-flip $[[3,2,1]]$ codes. The canonical CPC code ordering stipulates that phase-checks are performed before the bit-checks. Under this ordering, and by applying the previously described error propagation rules, it can be verified that the code in Figure \ref{fig:422_code}a will detect single-qubit $X$- and $Z$-errors that occur on either of the data qubits. However, as shown by the arrows in Figure \ref{fig:422_code}a, a phase-flip error on the parity qubit connected to the data qubits via bit-check edges propagates errors to the register that go undetected. As a result, the code depicted in Figure \ref{fig:422_code}a is a $[[4,2,1]]$ code. However, the undetected error propagation pathway can be closed by applying a cross-check edge between the two parity qubits as shown in Figure \ref{fig:422_code}b. This cross-check ensures that the code  detects all single-qubit $X$- and $Z$- errors, and so fixes the code distance to $d=2$. The code shown in Figure \ref{fig:422_code}b is therefore a $[[4,2,2]]$ code, capable of detecting both bit- and phase-flip errors on any of the four qubits.

\subsection{Annotated operational representation for general CPC codes} \label{sec:virtual_edges}

All CPC codes follow the same general construction as the $[[4,2,2]]$ code. Two codes -- one for bit-flips and one for phase-flips -- are merged to form a combined code. The code distance is then fixed via the addition of cross-checks between the parity qubits. Under the canonical CPC ordering, the phase-checks are performed first, the bit-checks second and the cross-checks last. Any CPC code can be represented via the operational representation consisting only of the qubit nodes and edges described in Section \ref{sec:422_code}. For larger codes, it is often useful to annotate the operational representation graph to better visualise the pathways for indirect propagation of errors. As an example, consider the case of a data qubit connected to two parity qubits, the first via a bit-check edge (red) and the second via a phase-check edge (black). A $Z$-error that occurs on the parity qubit to the right will be copied to the data qubit as an $X$-error when the phase-check operation is applied. The resultant $X$-error on the data qubit is then propagated to the other parity qubit by the bit-check operation. At the end of the CPC cycle, the initial $Z$-error is therefore detected as a bit-flip error on the other parity qubit. We can annotate the operational representation as follows:
\beq
\scalebox{0.7}{\input{figures/eq6.tikz}}.
\eeq
Here the error propagation pathway between the two parity qubits is highlighted by a directed pink edge. This edge can be considered a \emph{virtual edge}, as it does not correspond to a physical operation between the qubits it connects. The process of adding the virtual edges to an operational representation is called annotation. The net effect of the virtual edge in terms of error detection is therefore that a $Z$-error on the data qubit at the root of the arrow is propagated as an $X$-error on the qubit at the head of the arrow.

In addition to the case described above, there is another pathway for the indirect propagation of an error through a CPC code. Consider a data qubit connected to a parity qubit by both a bit-check and a phase-check edge. A phase-flip error on the parity qubit will propagate to the register and then back to the parity qubit as a bit-flip error. The annotation works as follows:
\beq
\scalebox{0.8}{{\centering \input{figures/eq7.tikz}}}
\eeq
This propagation can be described in terms of a self-loop virtual edge. The annotation is performed by adding exactly one virtual edge for each pair of bit and phase check edges incident on the same data qubit. Once all of these pairs have been considered, the process is complete.

In larger CPC codes, there will be multiple virtual edges that can cancel each other out. The addition of virtual edges can also lead to simplifications that reduce the total number of phase- and bit-check edges. A complete list of rules for adding virtual edges, along with various simplification rules, is included in Table \ref{tab:simp_rules} in Appendix \ref{app:qfg_prop_rules}. Formally the simplification process is the process of cancelling redundant propagation and therefore expressing the annotated factor graph in the simplest possible form. The process of simplification does not change which errors are propagated to where, it just refines this information to the most compact possible form.

The advantage of the annotated operational representation is that the virtual edges provide a way of illustrating the propagation of errors without having to consider the canonical CPC ordering. We will see in the following section that this simplification helps with the mapping from the operational representation to classical factor graphs.

\section{Mapping the graphical language to graphical models in classical information theory}\label{sec:qfg_to_cfg}

\subsection{Translation rules mapping the operational representation to classical factor graphs}

The graphical language of the operational representation, outlined in the previous section, allows quantum codes to be illustrated in terms of the physical operations connecting qubits. The annotated version of a operational representation includes virtual edges that highlight indirect propagation pathways for errors. We now show how the annotated operational representation can be mapped to an equivalent classical factor graph notation.

The data and parity nodes in the operational representation correspond to qubits that store both bit and phase error information. In a classical factor graph, a qubit can therefore be represented as two bits, one for each type of error. 
A data qubit is mapped to two classical data bits, one representing the bit information and the other the phase information:
\beq
{\centering\scalebox{0.8}{\input{figures/eq8.tikz}.}}
\eeq
As a convention, we choose to draw these bits side-by-side, with the node representing bit information on the left (coloured yellow) and the node representing phase information on the right (coloured blue).
A parity qubit is also mapped to two bits in classical factor graph notation: 
\beq
{\centering\scalebox{0.8}{\input{figures/eq9.tikz}.}}
\eeq
The bit information component of a qubit is used as a parity measurement, and is therefore drawn as a classical parity check node. The phase information of parity check qubit, however, cannot be directly measured. As such, the phase-information component of the parity check qubit is mapped to an unmeasured classical data bit (shown in blue on the right).

We can now describe how the different edge types in the operational representation are drawn in a classical factor graph. Bit-check edges connect data qubits to parity qubits. Their action is to propagate bit information from the data qubit to the parity qubit and phase information in the opposite direction. The mapping of a bit-check edge to classical factor graph notation is shown here:
\beq
{\centering \scalebox{0.8}{\input{figures/eq10.tikz}}}. 
\eeq
An edge is drawn between the bit information component of the data qubit and the bit information component of the parity qubit. Notice, however, that there is no edge drawn between the phase-components of two qubits to indicate the propagation of phase-flip errors from the parity qubit to the data qubit. This is omitted, as there is no concept of indirect error propagation in a classical factor graph; the edges in a classical factor graph are only permitted between data and parity nodes, and not between nodes of the same type. Instead, indirect propagation of errors are accounted for in classical factor graphs by placing edges in the place of the virtual edges in an annotated operational factor graph. An explicit example of this is shown later in this section.

The classical factor graph representation of a phase-check edge reads:
\beq
{\centering\scalebox{0.8}{\input{figures/eq11.tikz}}}. 
\eeq
The phase component of the data qubit is connected to the bit-component of the parity qubit via an edge. This shows that the phase-check edge propagates phase-errors on data qubits to bit-errors on the parity qubits. Recall that phase-edges are symmetric and that error propagation also occurs in the reverse direction. The back-propagation of errors in this way is not shown in the classical factor graph. The reason for this is again that edges are only permitted between data and parity nodes in a classical factor graph.

The classical factor graph representation of a cross-check edge reads: 
\beq
{\centering\scalebox{0.8}{\input{figures/eq12.tikz}}}. 
\eeq
The phase-component of each qubit is connected to the bit-component of the other. This reflects the expected error propagation behaviour for cross-check edges.

The next component to be mapped to classical factor graphs are the virtual edges that depict the indirect propagation of errors through the code. Virtual edges show how a phase-error on one parity qubit can be detected as a bit-flip error on another. The classical factor graph mapping of a virtual edge is given by
\beq
{\centering\scalebox{0.8}{\input{figures/eq13.tikz}. }}
\eeq
A virtual edge is directed meaning error information only propagates in one direction.

The final translation rule is for the virtual self loop edge. In the classical factor graph notation this edge is represented as follows:
\beq
{\centering\scalebox{0.8}{\input{figures/eq14.tikz}.}}
\eeq
A summary of the translation rules for mapping the operational representation to the classical factor graph representation can be found in Appendix \ref{app:translation_rules}.

\subsection{Translation example: The $[[4,2,2]]$ detection code} \label{ref:422translation}

We have now outlined how to translate the operational representation to classical factor graphs. Here we provide an example by analysing the development of the $[[4,2,2]]$ detection code in terms of both representations. 

In Section \ref{sec:422_code} we showed how a preliminary CPC code can be constructed by combining the bit-flip and phase-flip $[[3,2,1]]$ codes. The distance of this code was determined to be $d=1$ by consideration of the propagation of bit- and phase-flip errors through the code.  The code distance was then fixed to desired length of $d=2$ via the addition of a cross-check edge between the two parity qubits. The $[[4,2,2]]$ code is the simplest CPC detection code, and as such the code distance can essentially be determined by inspection. However, for larger CPC codes, calculating the code distance may become a hard problem. In these cases, however, other quantities such as error rate measured by Monte Carlo could be used as a metric of code performance instead \cite{Davey98montecarlo}.

\begin{figure}
	\centering
	\scalebox{0.8}{\input{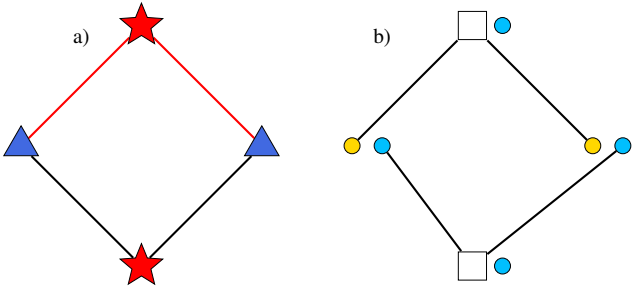}}
	\caption{The $[[4,2,1]]$ CPC code formed by combining the bit-flip and phase-flip $[[3,2,1]]$ codes. (a) The annotated operational representation. The virtual edges connect the same nodes in the same direction. They therefore cancel each other out, as per the rules outlined in Appendix \ref{app:qfg_prop_rules}. (b) The code in classical factor graph form. It can immediately be seen that there are two unconnected data bits, meaning the code has distance $d=1$.}
	\label{fig:421_to_classical}
\end{figure}

\begin{figure}
	\centering
	\scalebox{0.8}{\input{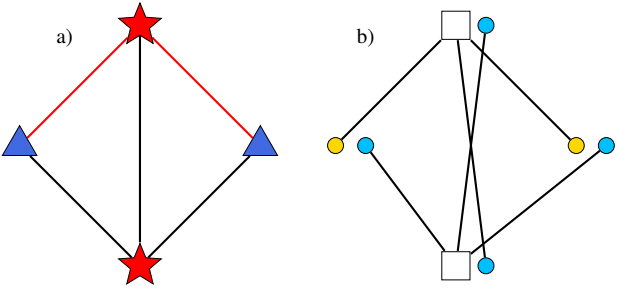}}
	\caption{The $[[4,2,2]]$ detection code. (a) The operational representation. (b) Classical factor graph version of the code. The edges that are added to the classical factor graph correspond to a cross-check edge in the operational representation.}
	\label{fig:422_to_classical}
\end{figure}

Figure \ref{fig:421_to_classical}a shows the annotated operational factor graph for the preliminary code formed by combining the bit-flip and phase-flip $[[3,2,1]]$ codes. As both of the virtual edges connect the same nodes in the same direction they cancel each other out, in accordance with the simplification rules outlined in Appendix \ref{app:qfg_prop_rules}. By following the mapping procedure outlined in the previous subsection, we obtain the classical factor graph of the preliminary code which is shown in Figure \ref{fig:421_to_classical}b. Inspection of the classical factor graph reveals that there are two data bits that go unchecked, from which it can be concluded that the distance of the code is $d=1$. The classical distance of the code represented by the classical factor graph is the same as the quantum distance of the operational factor graph it is based on. The preliminary CPC code is therefore a $[[4,2,1]]$ code. We discussed in Section \ref{sec:422_code} how to move from the $[[4,2,1]]$ to the $[[4,2,2]]$ code by the addition of a cross-check edge between the two parity qubits. The corresponding annotated operational representation is shown in Figure \ref{fig:422_to_classical}a which translates into the classical factor graph shown in Figure \ref{fig:422_to_classical}b.

\section{Designing quantum error correction codes without knowing quantum mechanics}

\subsection{General design rules to develop quantum error correction codes using classical factor graphs}

The factor graph formalism provides a highly general tool for the design of quantum error correction codes. We now outline a specific code design strategy that can be employed using classical factor graphs, without having to refer back to the operational form after the initial mapping. Our approach enables quantum codes to be constructed using classical techniques and does not require detailed knowledge of quantum mechanics. The steps of this code design procedure can be summarised as follows:
\begin{enumerate}
	\item Construct a preliminary code in the annotated operational representation by combining two classical codes, the first for bit-flip errors and the second for phase-flips. Convert the resultant graph to a classical factor graph using the mappings described in Section \ref{sec:qfg_to_cfg}. 
	\item Calculate the distance of the preliminary code. If code distance is not tractable, then use another metric such as simulated error rate.
	\item Determine the form of the cross-checks that need to be added to fix the code distance to the desired length. If code distance is not tractable, then use another metric such as simulated error rate.

\end{enumerate}
Following the initial mapping from the operational representation, the optimisation steps of the code design process (steps 2 and 3) are carried out entirely within the classical factor graph framework. The reason that this method can be followed without reference to the operational form is that the addition of cross-checks does not lead to any indirect propagation of errors. In a classical factor graph, cross checks are added between parity qubits according to the following rule,

\beq
{\centering\scalebox{0.8}{\input{figures/eq15.tikz}}},
\eeq
where a pair of edges link the phase component of one qubit to the bit component of the other. A situation which may be encountered when using this rule, occurs when a cross-check is applied between qubits that are already connected by one or more edges. For this case, we need to define a simplification rule for a double edge. Since the {\sc xor} of a bit value with itself is always zero, it stands to reason that two edges between the same pair of nodes in a factor graph will cancel as follows,
\beq
{\centering\scalebox{0.8}{\input{figures/eq16.tikz}.}}
\eeq
As an example, consider the case, depicted below, in which a cross-check is added between a pair of qubits that are already connected by a virtual edge, 
\beq
{\centering\scalebox{0.8}{\input{figures/eq17.tikz}.}}
\eeq
The double edge that is formed cancels to give the factor graph on the right. From this, we can deduce the rule that when a cross-check is added between a pair of qubits already connected by a virtual edge, the direction of the virtual edge is reversed. The operational version of this rule is listed in Table \ref{tab:simp_rules} in Appendix \ref{app:qfg_prop_rules}.

\subsection{Example: Designing a quantum error correction code based on the Hamming code} \label{sec:10_4_3}

\begin{figure*}[]
	\centering
	\scalebox{0.6}{\input{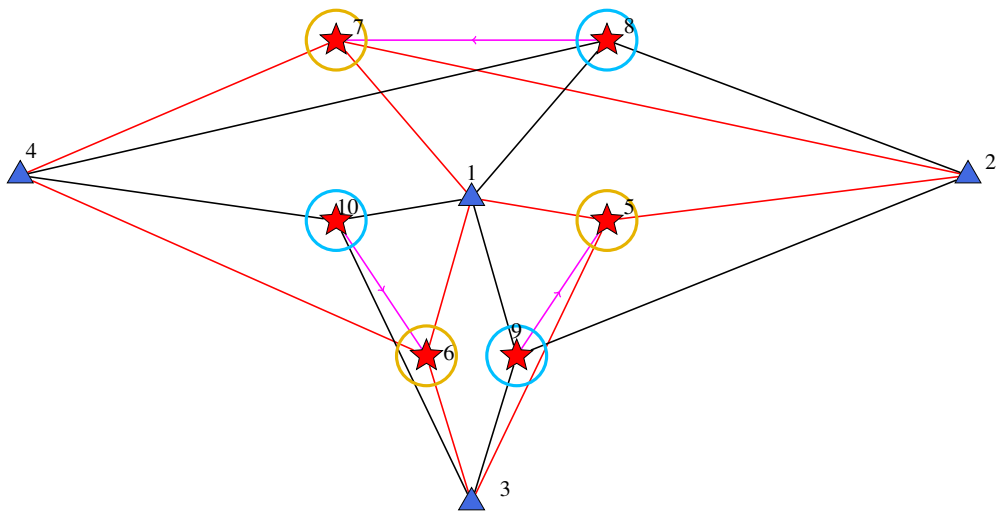}}
	\caption{The annotated operational representation of the preliminary code formed by combing two copies of the classical Hamming code, the first to detect bit-flips and the second to detect phase-flips. The parity qubits that detect bit-errors on the data qubits are circled in yellow, and the parity qubits that detect phase-flips on the data qubits are circled in blue. The virtual edges (pink) are added to the graph following the rules outlined in Table \ref{tab:simp_rules}. The qubits are numbered 1-10.}
	\label{fig:10_4_3_operational}
\end{figure*}

\begin{figure*}
	\centering
	\scalebox{.6}{\input{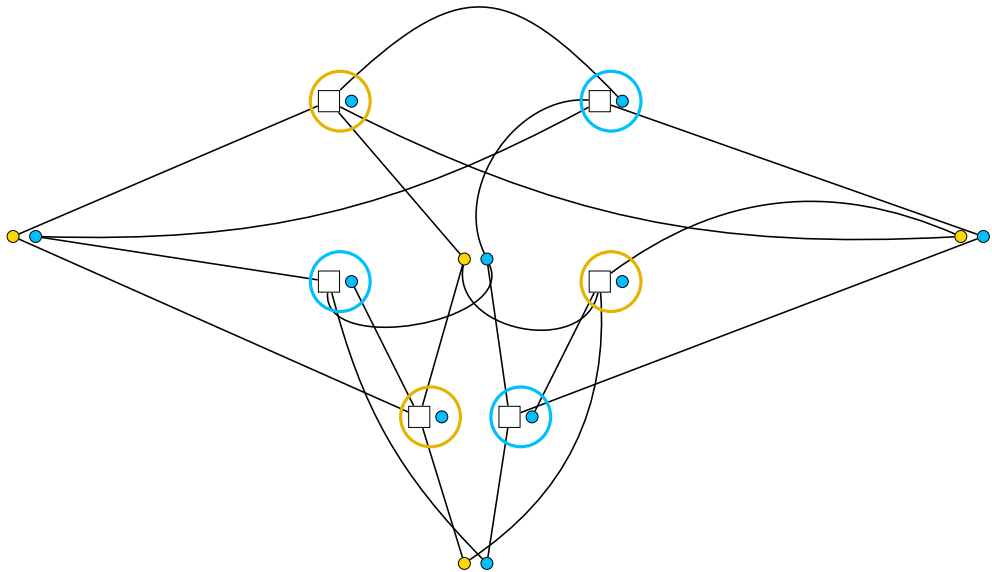}}
	\caption{The classical factor graph for the preliminary code formed by combining two copies of the classical Hamming code depicted in Figure \ref{fig:cfg1}b. Parity check qubits belonging to $S_1$ are circled in yellow, while those belonging to $S_2$ are circled in blue. The phase-components of each parity check qubit are not connected to any other nodes. As a result, $Z$-errors of the parity check qubits go undetected, meaning the code has distance $d=1$. The preliminary code is therefore a $[[10,4,1]]$ code. The operational representation of this code is shown in Figure \ref{fig:10_4_3_operational}.}
	\label{fig:two_7_4_3_classical}
\end{figure*}

We now apply the code design method outlined in the previous subsection to a small example. We first combine two copies of the classical Hamming code given in Figure \ref{fig:cfg1}c to form a preliminary quantum code. After applying the indirect error propagation annotations (following the rules given in Table \ref{tab:simp_rules} in appendix \ref{app:qfg_prop_rules}), the operational representation of the preliminary code is given by Figure \ref{fig:10_4_3_operational}. This annotated operational form of the preliminary code then maps to the factor graph depicted in Figure \ref{fig:two_7_4_3_classical}. From this point onwards, the construction and optimisation of the code proceeds entirely within the classical factor graph framework. By inspection, we see immediately that the preliminary code has some nodes which are not joined to any parity check nodes. These nodes correspond to the phase-components of the parity check qubits. As a result, some errors will go undetected, meaning the preliminary code is a $[[10,4,1]]$ quantum code. 

To turn the preliminary code into a distance three quantum error correcting code, each data node must be connected to at least two parity checks. We must further ensure that all single-bit errors result in a unique error pattern. By construction, the parity check qubits in the code in Figure \ref{fig:two_7_4_3_classical} fall into two disjoint subsets. 
\begin{itemize}
	\item Subset $S_1$: the parity check qubits which detect $X$-errors (bit flip) on data qubits (circled in yellow).
	\item Subset $S_2$: the parity check qubits which detect $Z$-errors (phase) on data qubits (circled in blue).
\end{itemize}
This separation will help in categorising the different types of errors and the error patterns they produce. For example, it can be seen that bit-flip errors on the data qubits will result in an error pattern involving only parity check qubits in $S_1$. Similarly, phase-flip errors on the data qubits result in an error pattern involving only the parity check qubits in $S_2$.

\begin{figure*}
	\centering
	\scalebox{.6}{\input{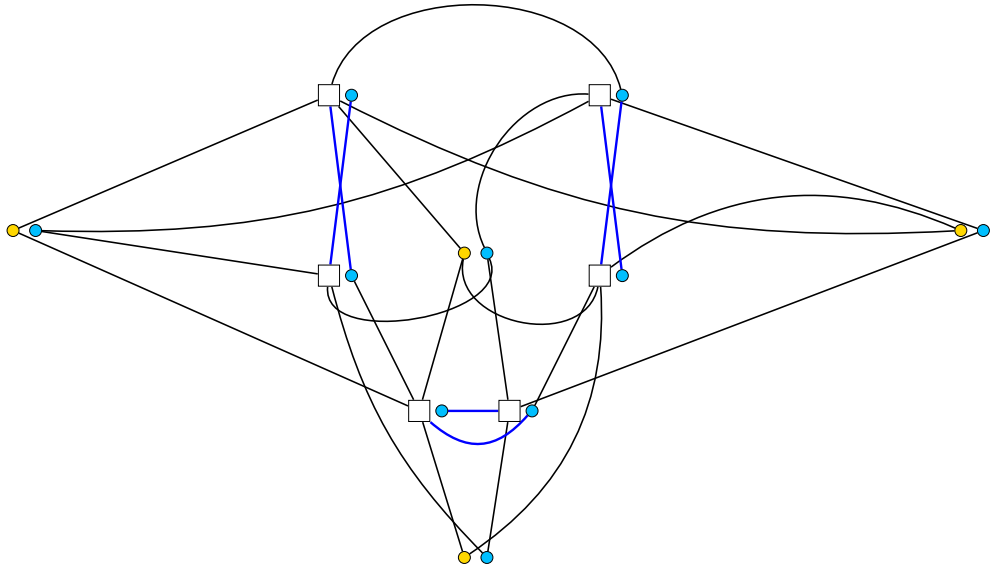}}
	\caption{The code formed by the addition of cross-check edges (highlighted in blue) between parity check qubits in $S_1$ and $S_2$. As each bit node in the graph is connected to least one parity check node, this is a $[[10,4,2]]$ detection code.}
	\label{fig:10_4_3_classical_stage1}
\end{figure*}

The first modification to be made to the preliminary code is to add edges to the unconnected phase nodes in each parity check qubit. This can be achieved through the addition of cross-checks between the affected parity check qubits, as depicted by the blue edges in Figure \ref{fig:10_4_3_classical_stage1}. Notice that we have specifically chosen cross-checks that connect parity qubits in $S_1$ to parity qubits in $S_2$. This ensures that the error patterns resulting from $Z$-errors on the parity check qubits will contain a combination of measurements from $S_1$ and $S_2$. Consequently, $Z$-errors on the parity check qubits are distinguishable from $X$- and $Z$-errors on the data bits. As there are no longer are any unconnected bits, the modified code in Figure \ref{fig:10_4_3_classical_stage1} is a $[[10,4,2]]$ detection code.

\begin{figure*}
	\centering
	\scalebox{.6}{\input{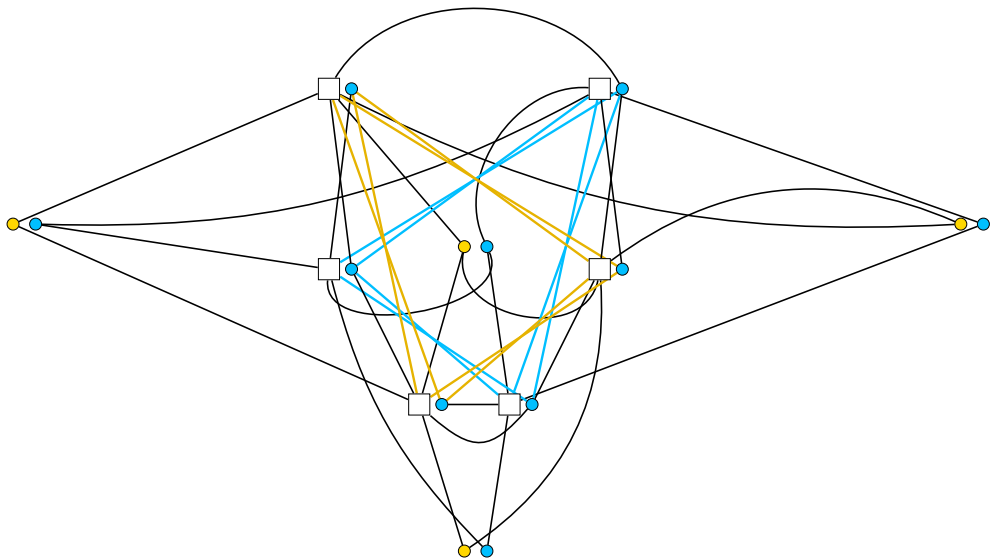}}
	\caption{The code formed by the addition of cross-checks between all the qubits in $S_1$ and likewise for $S_2$. The yellow cross-checks connect qubits in $S_1$, and the blue cross-checks connect qubits in $S_2$. This code will produce unique syndromes for an error model containing $X$-, $Z$- and $Y$-errors. It is therefore a $[[10,4,3]]$ code.}
	\label{fig:10_4_3_classical_stage2}
\end{figure*}

We now have a code that will detect $Z$-errors on any of the parity qubits. The next step is to find a code modification that will guarantee that the error patterns produced by these errors are distinguishable. One way of achieving this is to add cross-checks between all parity qubits in $S_1$ and likewise for $S_2$. The final factor graph is shown in Figure \ref{fig:10_4_3_classical_stage2}, where the cross-checks connecting $S_1$ are shown in yellow and the cross-checks connecting $S_2$ are shown in blue. Based on previous arguments, this code is now a $[[10,4,3]]$ code for an error model containing $X$- and $Z$-errors.

We now show that the $[[10,4,3]]$ code depicted in Figure \ref{fig:10_4_3_classical_stage2} will also produces unique error patterns for another type of error, the $Y$-error. In terms of the classical factor graphs, $Y$-errors correspond to a specific type of burst error in which both the bit- and phase-components of a qubit are errored simultaneously. We first note that all $Y$-errors on data qubits result in detection patterns of weight four or six with half of the parity measurements in $S_1$ and half in $S_2$. Fortunately, the error patterns with two parity measurement in each subset are distinct from those produced by a $Z$-error in $S_2$. On the other hand, $Y$-errors on $S_1$ parity check qubits will result in error patterns of weight four, of which three of the measurements occur in $S_1$ and one of the measurement in $S_2$. For $Y$-errors on $S_2$, there will be three measurements in $S_2$ and two in $S_1$. Since the weight of the error patterns resulting from $Y$-errors are greater than that for $X$ or $Z$, the signatures that are produced are unique. The factor graph in Figure \ref{fig:10_4_3_classical_stage2} is therefore also a $[[10,4,3]]$ code for an error model containing $X$-, $Z$- and $Y$-errors. Table \ref{tab:S1_S2_num} summarises the structure of signatures produced for different types of errors in the $[[10,4,3]]$ code.

The complete syndrome table for the $[[10,4,3]]$ code shown in the factor graph in Figure \ref{fig:10_4_3_classical_stage2} is shown in Table \ref{tab:syndromes} in Appendix \ref{app:syndrome}. We have now shown that a distance $d=3$ quantum code can be constructed from a starting point of two classical codes using factor graph methods. The necessary modifications to the code were deduced entirely through visual inspection of the classical graph. Cross-checks were chosen in such a way that different types of error on different components of the code produced error signatures of distinct types. It should be noted that the solution we have found is not unique; other combinations of cross-checks exist that will also fix the code distance to $d=3$.

\renewcommand{\arraystretch}{1.5}
\begin{table}[t]
	\begin{centering}
	\caption{Table listing the structure and weight of the error signatures (also refereed to as syndromes) resulting from different types of errors in the $[[10,4,3]]$ code depicted in Figure \ref{fig:10_4_3_classical_stage2}. Each error-type results in a different syndrome structure, meaning single-qubit errors are distinguishable. The full syndrome list for this code can be found in Appendix \ref{app:syndrome}.\label{tab:S1_S2_num}}
		{
		\begin{tabular}{|c|c|c|}
			\hline 
			error type & \# of detections in $S_{1}$ & \# of detections in $S_{2}$\tabularnewline
			\hline 
			\hline
			$X$ data qubit & 2 or 3 & 0\tabularnewline
			\hline 
			$Z$ data qubit & 0 & 2 or 3\tabularnewline
			\hline 
			$Y$ data qubit & 2 if 2 in $S_{2}$ or 3 if 3 in $S_{2}$  & 2 or 3\tabularnewline
			\hline 
			$X$ , $S_{1}$ & 1 & 0\tabularnewline
			\hline 
			$Z$ , $S_{1}$ & 2 & 1\tabularnewline
			\hline 
			$Y$ , $S_{1}$ & 3 & 1\tabularnewline
			\hline 
			$X$ , $S_{2}$ & 0 & 1\tabularnewline
			\hline 
			$Z$ , $S_{2}$ & 2 & 2\tabularnewline
			\hline 
			$Y$ , $S_{2}$ & 2 & 3\tabularnewline
			\hline 
		\end{tabular}}
		\par\end{centering}
	\vspace*{3mm}
	
\end{table}

\section{Discussion}

We have introduced a framework to create classical graphical models or factor graphs for quantum error correcting codes. This process begins with an operational representation, designed to show how the qubits interact to form a coherent parity check code \cite{chancellor16}. The operational representation given here is a \emph{human readable} analogue of the \emph{machine readable} matrix based formalism in \cite{chancellor16}. Unlike the more general ZX calculus used in \cite{chancellor16}, this is a special-purpose graphical representation specifically designed to show error propagation. Because of the nature of the interactions between qubits, there is unavoidable indirect propagation of errors. However, this propagation can be understood in terms of graphical rules which amount to the addition of virtual edges to the operational representation. Once the virtual edges have been added to the operational representation, there is a further set of graphical rules to map it to a classical factor graph. This classical factor graph represents error propagation within the quantum code. 

We demonstrated a design procedure using classical factor graphs which requires no reference to the operational representation, and allows for quantum code design to be treated as classical problem with restrictions on how interactions may be added. This means that code design can be performed based on completely classical intuition about error correction. Highly-optimised tools for classical LDPC and turbo codes can therefore be applied to quantum error correction, without the user having to understand quantum theory. An interesting direction for future work would be to apply our CPC factor graph methods to existing constructions for converting LDPC codes to quantum codes \cite{Tillich2014}.

In the design example we have given here, we have calculated code distance to verify the performance of our code. For larger codes with larger code distances, such a calculation will not be possible, and performance would have to be verified by another method, for example, calculating logical error rates using Monte Carlo \cite{Davey98montecarlo}. The real power of our method is that it allows a visual representation of any (CPC formulated) quantum code which on one hand directly corresponds to the physical interactions between qubits through the operational representation, and on the other hand to the processing of error information through the factor graph representation. These graphical representations allow a very general handle for human intuition to be used in the design process without prior knowledge of quantum mechanics. The primary purpose of this paper is not to propose a specific new design technique for quantum codes (although we propose one to demonstrate how our techniques can be used), but to provide an additional tool which can be used either on its own or in conjunction with known techniques.

While it is likely that the graphical methods proposed here could be used on their own to produce highly competitive quantum codes, our design methods do not need to be used in a stand-alone setting. For instance if a good CSS code were already known, our methods could be used to make modifications to improve it, for instance by making a version which is more compatible with the allowed interactions in a given quantum hardware, or by performing local modifications to try to further reduce the logical error rate. The second of these would be particularly useful if errors on a subset of the physical qubits were identified as being disproportionately responsible for logical errors. Because a broad class of existing quantum codes can be represented in the CPC framework, the methods proposed here can be used to augment the existing toolkit of techniques for quantum error correction.

We have shown the utility of our graphical representation with both error detection codes and error correction codes. In general, factor graph methods can be applied equally well to codes that are designed to suppress errors rather than fully correct. Many of the most promising quantum algorithms for quantum simulation (i.e.~the variational quantum eigensolver \cite{kandala17a,moll17a,Wang18a}) and optimisation (i.e.~the quantum approximate optimisation algorithm \cite{Farhi14a,Farhi14b,Yang17a,Farhi17a}) are tolerant to some error. In this context, bespoke error suppression codes designed using factor graphs would be useful.  

A further advantage to the classical factor graph representation is that the graph effectively tracks all of the detected errors, and can therefore be viewed as a simplified simulation of the code's error propagation pathways. This means that the factor graph representation can reliably give information about error propagation of complex and correlated errors. To illustrate this, we consider the propagation of $Y$-errors which arise on quantum hardware when a bit- and phase-flip occur simultaneously. Such errors can be thought of as a two-bit burst error, occurring on both the bit and phase part of a qubit simultaneously. 

Although we do not provide an example in this paper, it is possible to construct CPC codes that include a \textit{second-tier} of parity qubits whose role it is to monitor other parity qubits. A construction of this type is outlined in \cite{chancellor16} to provide a general method for converting any pair of distance three classical codes to a quantum code. As the second-tier of parity qubits only check other parity bits, they do not interact with the data qubits. Because of this, errors that occur on second-tier parity qubits can be considered \emph{benign} in the sense that they do not propagate errors to the data qubits. As such, it is only necessary to detect the benign errors instead of fully correcting. Performance metrics, such as code distance, are calculated without taking into account benign errors, and can therefore be thought of as a lower bound on the code performance based on a conservative decoding strategy. The codes outlined in this paper do not include second-tier parity qubits, and benign errors do not need to be considered. However, in future work it would be interesting to search for codes where consideration of benign errors becomes important, and to modify our notation to account for this.     

An important consideration in the design of quantum codes is that some interactions between qubits may be difficult or impossible to implement directly on real quantum hardware. The fact that the physical layout of the device is important, highlights another strength of our graphical formalism: the graph created by the operational representation of the code is the interaction graph of the qubits. It is therefore natural to include hardware constraints into the methods given here. This allows for the design of quantum codes specially optimised to the demands of a given quantum device. 

The search for bespoke codes for given quantum hardware can be further assisted by machine learning techniques. The framework presented here is particularly powerful with regard to this. Firstly, we can use design patterns from classical codes to construct quantum codes with specific features. Secondly, we can start from known quantum codes and perform local search around those as well as impose (global) constraints on them as mentioned above. Modern machine learning algorithms combine techniques from Bayesian optimisation for global search with other local search methods, such as simulated annealing, and automatically switch between those. It is in this situation that the CPC construction in general, and the graphical methods given in particular, will be especially useful. The structural representation, and the ability to search over propagation of errors, makes it particularly amenable for use in small-change optimisation strategies. This means any given code (including e.g. one's favourite existing quantum or classical code) can be used as input and changed slightly in order to optimise it for specific hardware or architectural considerations. In future work we plan to combine the graphical framework presented in the work together with the above machine learning techniques to provide methods of automated code design.   

In summary, we have given in this paper a new way to connect the knowledge and skills of classical information processing to the design of quantum error correction procedures. By representing error propagation in an intuitive, graphical, and classical-style way, the problem of designing and simulating quantum codes becomes much more tractable.
The connection to classical graphical representations is both interesting theoretically and of powerful practical use.
The expectation is that these tools will allow the skills and intuition of the classical error correction community to be brought to bear on the next generation of quantum error correction codes.

\section*{Acknowledgements}

\begin{table*}[]
	
	\begin{centering}
		\caption{(Top block) The annotation rules for adding virtual edges in the operational representation of CPC codes. The annotation rules are applied once before any graph simplifications are applied. (Bottom block) Simplification rules for annotated factor graphs. These rules are applied repeatedly until no simplifications are possible. The order does not matter as long as the first block of annotation rules is applied first. \label{tab:simp_rules}}
		\def\arraystretch{5}
		\begin{tabular}{|c|c|c|}
			\hline 
			Rule & Before Rule & After Rule  \tabularnewline
			\hline 
			\hline 
			\begin{minipage}[t]{0.2 \textwidth} \vspace{- 1.0cm} Virtual \\ Edge \\  Creation \end{minipage} & \input{figures/pre_virt_edge.tikz} & \input{figures/post_virt_edge.tikz}\tabularnewline
			\hline 
			\begin{minipage}[t]{0.2 \textwidth} \vspace{- .8cm}  Virtual\\ Loop \\ Creation\end{minipage}  & \input{figures/pre_loop.tikz} & \input{figures/post_loop.tikz} \tabularnewline
			\hline
			\hline
			\begin{minipage}[t]{0.2 \textwidth} \vspace{- .50 cm}  Virtual Edge \\ Cancellation\end{minipage} & \input{figures/pre_virt_cancel.tikz}  & \input{figures/post_virt_cancel.tikz} \tabularnewline
			\hline 
			\begin{minipage}[t]{0.2 \textwidth} \vspace{- 1.15 cm}  Virtual\\ Loop \\ Cancellation\end{minipage} & \input{figures/pre_loop_cancel.tikz} & \begin{minipage}[t]{0.2 \textwidth} \vspace{-1.25 cm}\input{figures/post_loop_cancel.tikz} \end{minipage} \tabularnewline
			\hline 
			\begin{minipage}[t]{0.2 \textwidth} \vspace{- .55 cm}  Virtual Edge \\ Reversal\end{minipage} & \input{figures/pre_virt_reversal.tikz} & \input{figures/post_virt_reversal.tikz} \tabularnewline
			\hline 
			\begin{minipage}[t]{0.2 \textwidth} \vspace{- .55 cm}  Virtual Edge \\ Addition\end{minipage} & \input{figures/pre_virt_add.tikz} & \input{figures/post_virt_add.tikz} \tabularnewline
			\hline 
		\end{tabular}
		\par\end{centering}
	\vspace*{3mm}

\end{table*}

Joschka Roffe acknowledges funding from a Durham Doctoral Studentship (Faculty of Science) and the support of the QCDA project which has received funding from the QuantERA ERA-NET Cofund in Quantum Technologies implemented within the European Union’s Horizon 2020 Programme. Nicholas Chancellor acknowledges funding from EPSRC grant refs EP/L022303/1 and EP/S00114X/1. Dominic Horsman acknowledges funding from EPSRC grant ref EP/L022303/1 and  the “Investissements d’avenir” (ANR-15-IDEX-02) program of the French National Research Agency. Stefan Zohren is funded by the Oxford-Man Institute. The authors thank Aleks Kissinger and Viv Kendon for useful discussions. The diagrams in this paper were produced using the Tikzit tikz editor \cite{tikzit}. The authors also acknowledge the use of the NumPy Python package for calculations associated with this paper\cite{numpy}.

\begin{table*}[h!]
	\begin{centering}
		\caption{Rules for mapping the annotated operational representation to the classical factor graph representation. \label{tab:translation_rules}}
		\def\arraystretch{5}
		\begin{tabular}{|c|c|c|}
			\hline 
			Description & Operational representation & Classical Factor Graph \tabularnewline
			\hline 
			\hline 
			\begin{minipage}[t]{0.05 \textwidth} \vspace{- 1.2cm} Phase \\ Check \\  Edge \end{minipage} & \input{figures/phase_edge.tikz} & \input{figures/phase_edge_class.tikz}\tabularnewline
			\hline 
			\begin{minipage}[t]{0.05 \textwidth} \vspace{- 1.2cm}  Bit\\ Check \\Edge \end{minipage}  & \input{figures/bit_edge.tikz} & \input{figures/bit_edge_class.tikz} \tabularnewline
			\hline 
			\begin{minipage}[t]{0.05 \textwidth} \vspace{- 1.2cm} Cross \\ Check \\ Edge \end{minipage} & \input{figures/cross_check.tikz}  & \input{figures/cross_check_class.tikz} \tabularnewline
			\hline 
			\begin{minipage}[t]{0.05 \textwidth} \vspace{- 1.2cm}  Virtual \\Edge \end{minipage} & \input{figures/virt_edge.tikz} & \input{figures/virt_edge_class.tikz} \tabularnewline
			\hline 
			\begin{minipage}[t]{0.1 \textwidth} \vspace{- .7cm} Virtual \\ Self Loop \end{minipage} & \input{figures/virt_loop.tikz} &\input{figures/virt_loop_class.tikz} \tabularnewline
			\hline 
		\end{tabular}
		\par\end{centering}
	\vspace*{3mm}
\end{table*}

\appendices

\section{General rules for annotating graphs of the operational representation} \label{app:qfg_prop_rules}

The operational representation of a factor graph is annotated through the addition of directed virtual edges. These virtual edges show where $Z$-errors on the parity bits are detected. The rules for annotating operational factor graphs are shown in Table \ref{tab:simp_rules}. Table \ref{tab:simp_rules} also lists simplification rules that arise when multiple edges combine.

\section{General rules for annotating graphs of the operational representation} \label{app:qfg_prop_rules}

The operational representation of a factor graph is annotated through the addition of directed virtual edges. These virtual edges show where $Z$-errors on the parity bits are detected. The rules for annotating operational factor graphs are shown in Table \ref{tab:simp_rules}. Table \ref{tab:simp_rules} also lists simplification rules that arise when multiple edges combine.     

\section{General rules for mapping the operational representation to the classical factor graph representation} \label{app:translation_rules}

The rules for mapping the annotated operational graph to a classical factor graph are listed in Table \ref{tab:translation_rules}.

\begin{table*}[h!]
	\centering
		\caption{Syndrome table for the $[[10,4,3]]$ Hamming code depicted in operational form in Figure \ref{fig:10_4_3_operational} and in classical factor graph form in Figure \ref{fig:10_4_3_classical_stage2}. The syndromes for each error are represented as six-bit binary strings. The bits, from left-to-right, correspond to the measurement outcomes of the parity qubits $5-10$ (labelled in Figure \ref{fig:10_4_3_operational}). The first three bits in the syndrome string (coloured red) represent measurement outcomes from the parity qubits in subset $S_1$. The final three-bits (coloured blue) in the syndrome represent measurement outcomes of the parity-qubits in subset $S_2$.	\label{tab:syndromes}}  
	\label{my-label}
	\begin{tabular}{|l|l|c|c|c|}
		\hline
		&          & \multicolumn{1}{l|}{$X$-error syndrome} & \multicolumn{1}{l|}{$Z$-error syndrome} & \multicolumn{1}{l|}{$Y$-error syndrome} \\ \hline \hline
		\multicolumn{1}{|c|}{\multirow{4}{*}{\begin{tabular}[c]{@{}c@{}}data\\ qubits\end{tabular}}} 
		& qubit 1 & \synd{111}{000} & \synd{000}{111} & \synd{111}{111}                                         \\ \cline{2-5} 
		\multicolumn{1}{|c|}{}                                                                       
		& qubit 2 & \synd{101}{000} & \synd{000}{110} & \synd{101}{110}                                   \\ \cline{2-5} 
		\multicolumn{1}{|c|}{}                                                                       
		& qubit 3 & \synd{110}{000} & \synd{000}{011} & \synd{110}{011}                             \\ \cline{2-5} 
		\multicolumn{1}{|c|}{}                                                                       
		& qubit 4 & \synd{011}{000} & \synd{000}{101} & \synd{011}{101}                                      \\ \hline \hline
		\multirow{6}{*}{\begin{tabular}[c]{@{}l@{}}parity\\ qubits\end{tabular}}                     
		& qubit 5 & \synd{100}{000} & \synd{011}{100} & \synd{111}{100}                                      \\ \cline{2-5} 
		& qubit 6 & \synd{010}{000} & \synd{101}{010} & \synd{111}{010}                                  \\ \cline{2-5} 
		& qubit 7 & \synd{001}{000} & \synd{110}{001} & \synd{111}{001}                                                 \\ \cline{2-5} 
		& qubit 8 & \synd{000}{100} & \synd{101}{011} & \synd{101}{111}                                          \\ \cline{2-5} 
		& qubit 9 & \synd{000}{010} & \synd{110}{101} & \synd{110}{111}                                        \\ \cline{2-5} 
		& qubit 10 & \synd{000}{001} & \synd{011}{110} & \synd{011}{111}                                        \\ \hline
	\end{tabular}
	\vspace*{3mm}
	
\end{table*}

\section{Syndrome table for the $[[10,4,3]]$ code} \label{app:syndrome}

Table \ref{tab:syndromes} is the syndrome table for the $[[10,4,3]]$ code depicted in the classical factor graph in Figure \ref{fig:10_4_3_classical_stage2}. The qubit numbers correspond to the qubits labels shown in the operational form of the code in Figure \ref{fig:10_4_3_operational}. The syndromes can be inferred by direct inspection of the factor graphs. Alternatively, the syndrome table can be calculated using the matrix methods outlined in \cite{chancellor16,roffe17}. 

\section{The quantum circuit-model representation of CPC codes} \label{app:qcm}

\begin{figure}
	\centering
\input{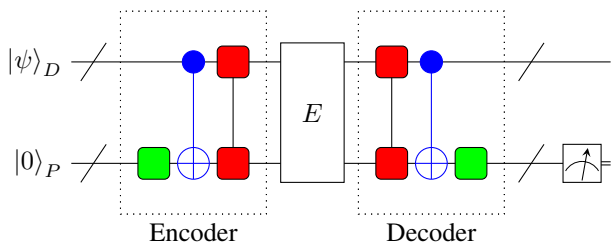}
\caption{The CPC code structure in circuit-model notation. The code qubits are partitioned into data qubits $\ket{\psi}_D$ and parity qubits $\ket{0}_P$. The encoder/decoder is split into three stages: 1) Cross-check stage (depicted by the green gate) between parity qubits; 2) Phase-check stage (depicted by the blue gate); 3) Bit-check stage (depicted by the red gate). The ordering of the three stages of the encoder is important, as it influences the derivation of rules for dealing with indirect propagation of errors in our graphical models. We refer to the ordering depicted here as the canonical ordering.}
\label{fig:can_form}
\end{figure}

The Coherent Parity Check (CPC) framework was first introduced in \cite{chancellor16} with the aim of providing new perspectives and tools for the construction of stabilizer codes. In this paper, the goal is to provide an introduction to the CPC framework exclusively in terms of a graphical representation that does not require prior knowledge of quantum circuit notation. For completeness, in this appendix, we provide a brief outline of the CPC framework in the quantum circuit picture. Prior knowledge of standard quantum circuit notation, as seen for example in \cite{Nielsen}, is assumed. A more thorough introduction to the CPC framework using quantum circuit notation can be found in \cite{roffe17}.

All CPC codes have a symmetric structure of the form shown in Figure \ref{fig:can_form}. The code qubits are separated into two distinct types, corresponding to the data register $\ket{\psi}_D$ and the parity register $\ket{0}_P$. The decoder of the circuit is split into three parts. In the first stage (shown by the red square gate in Figure \ref{fig:can_form}), a round of phase-checks are performed between the data register and the parity register. In the second stage (depicted by the blue gate in Figure \ref{fig:can_form}), a round of bit-checks are performed. The final stage of the CPC decoder involves performing a round of cross-checks between the parity qubits themselves (depicted by the green square in Figure \ref{fig:can_form}). Each of these check stages can take the parity checking sequence from an existing classical code. The encoder is simply the unitary inverse of the decoder. For well chosen parity checks, it is possible to detect errors that occur in the region marked $E$ by measuring the parity qubits at the end of the circuit.

Note that the ordering of the different checks in the encoder/decoder of the CPC circuit in Figure \ref{fig:can_form} is important, as it will influence the definition of graphical rules for indirect propagation. We refer to the specific ordering depicted in Figure \ref{fig:can_form} -- cross-checks, bit-checks and then phase-checks -- as the canonical form for CPC codes.  

We now describe the gates with which the three stages of a CPC circuit of the form depicted Figure \ref{fig:can_form} are realised. In addition, we provide a mapping for each circuit-model gate to the operational representation. Bit-checks (depicted by the blue gate in Figure \ref{fig:can_form}) are performed via {\sc{CNOT}} gates. A simple bit-check CPC gadget, with one data qubit and one parity qubit, is shown below
\begin{equation} \label{eq:op_rep_bit}
\vcenter{\hbox{
		\resizebox{!}{!}{\input{figures/eq18a.tikz}}}} \quad \longrightarrow \quad \vcenter{\hbox{ \resizebox{!}{20mm}{	\begin{tikzpicture}[style={data_qubit}]
	\begin{pgfonlayer}{nodelayer}
		\node [style={data_qubit}] (0) at (0, 1.5) {};
		\node [style={parity_check_qubit}] (1) at (0, -0) {};
	\end{pgfonlayer}
	\begin{pgfonlayer}{edgelayer}
		\draw [style={CNOT}] (0) to (1);
	\end{pgfonlayer}
\end{tikzpicture}}   }}\rm ,
\end{equation}
where the diagram to the right is the corresponding CPC circuit in the operational representation. Similarly, the phase-check stage involves conjugate-propagator gates. The simplest phase-check CPC gadget is given by
\begin{equation} \label{eq:op_rep_phase}
\vcenter{\hbox{
		\resizebox{!}{!}{\input{figures/eq19a.tikz}}}} \quad \longrightarrow \quad \vcenter{\hbox{ \resizebox{!}{20mm}{	\begin{tikzpicture}[style={data_qubit}]
	\begin{pgfonlayer}{nodelayer}
		\node [style={data_qubit}] (0) at (0, 1.5) {};
		\node [style={parity_check_qubit}] (1) at (0, -0) {};
	\end{pgfonlayer}
	\begin{pgfonlayer}{edgelayer}
		\draw [style={conj_prop}] (0) to (1);
	\end{pgfonlayer}
\end{tikzpicture}}   }}\rm ,
\end{equation}
where the diagram to the right is the operational representation of the circuit. The two-qubit gates with the black squares are the conjugate-propagator gates, which can be defined in terms of {\sc CNOT} gates as follows
\begin{equation}
\vcenter{\hbox{\input{figures/eq20.tikz}}}\rm .
\end{equation}
Finally, cross-checks are realized as conjugate-propagator gates acting between parity qubits. A simple example, and its mapping to the operational representation, is shown below
\begin{equation}\label{eq:op_rep_cross}
\vcenter{\hbox{
		\resizebox{!}{!}{\input{figures/eq21a.tikz}}}} \quad \longrightarrow \quad \vcenter{\hbox{ \resizebox{!}{20mm}{	\begin{tikzpicture}[style={data_qubit}]
	\begin{pgfonlayer}{nodelayer}
		\node [style={parity_check_qubit}] (0) at (0, 1.5) {};
		\node [style={parity_check_qubit}] (1) at (0, -0) {};
	\end{pgfonlayer}
	\begin{pgfonlayer}{edgelayer}
		\draw [style={conj_prop}] (0) to (1);
	\end{pgfonlayer}
\end{tikzpicture}}   }}\rm .
\end{equation}

We have defined the CPC codes such that in the encoder the bit information is propagated to the parity check qubits via CNOTs before the phase information is propagated. The decoder then performs the operations in reverse ordering. Fig. \ref{fig:can_form} depicts an example of this canonical ordering. Note that the cross check operations commute with both the bit and phase checks as well as each other: the time at which they are performed therefore does not affect the unitary implemented by the encoder and decoder. We however conventionally choose to draw the bit checks in the encoder.

As explained in Section \ref{sec:virtual_edges}, virtual edges depict how $Z$-errors on the parity qubits can be detected via indirect propagation pathways. We now illustrate such a pathway in the quantum circuit for a CPC code. Consider the circuit shown below
\begin{equation}\label{eq:op_rep_virtual1}
\resizebox{\columnwidth}{!}{$
\vcenter{\hbox{
		\resizebox{!}{!}{\includegraphics{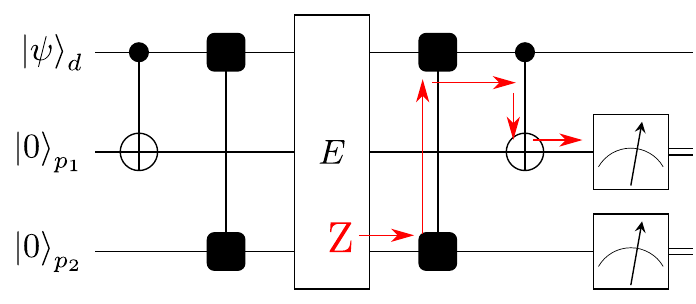}}}} \quad \longrightarrow \quad \vcenter{\hbox{ \resizebox{!}{20mm}{	\input{figures/eq22b.tikz}}   }}\rm . $}
\end{equation}
A data qubit $\ket{\psi}_D$ is connected to one parity via a {\sc CNOT} gate and to another via a conjugate-propagator gate. Under this setup, a $Z$-error in the wait stage on parity qubit $p_2$ will be propagated to the data register via the conjugate-propagator gate, before being propagated to parity qubit $p_1$ by the {\sc{CNOT}} gate (for methods to calculate propagation of this type see \cite{roffe17}). A $Z$-error on parity qubit $p_2$ is therefore detected indirectly on parity qubit $p_1$. In the operational representation, shown to the right of Figure \ref{eq:op_rep_virtual1}, this indirect propagation is depicted by the pink virtual edge.

\section{The quantum parity check matrix of a CPC code}
\begin{figure}
    \centering
    \includegraphics{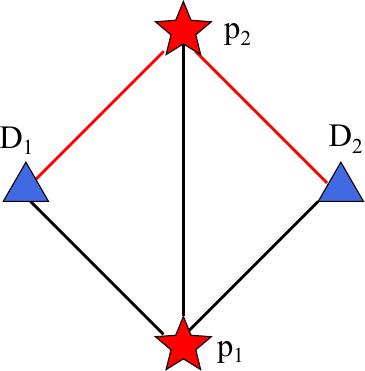}
    \caption{The $[[4,2,2]]$ CPC code in the operational representation.}
    \label{fig:422_app}
\end{figure}

This appendix describes how to deduce the stabilizers of a CPC code from the operational representation \cite{chancellor16,roffe17,Roffe_thesis}. As outlined in appendix \ref{app:qcm}, the encoder of a CPC code consists of three stages: cross-checks, bit-checks and phase-checks. Each of these stages can be thought of as a sub-graph of the operational representation with a corresponding adjacency matrix. In general, the adjacency matrices for a CPC code are given by
\begin{equation}
\resizebox{\columnwidth}{!}{$
\begin{split}
&m_b=\begin{blockarray}{cccccc}
\ & \text{\scriptsize $[p_1]$} & \text{\scriptsize $[p_2]$} & \text{\scriptsize [...]}  & \text{\scriptsize $[p_m]$}\\
\begin{block}{c(ccccc)}
\text{\scriptsize{$[D_1]$}} & b_{11} & b_{12} & ... & b_{1} \\
\text{\scriptsize{$[D_2]$}} & b_{21} & b_{22} & ... & b_{2m} \\
\text{\scriptsize{$[...]$}} & ... & ... & ... & ... \\
\text{\scriptsize{$[D_k]$}} & b_{k1} & b_{k2} & ... & b_{km} \\
\end{block}
\end{blockarray} \ \rm,
\\
&m_p=\begin{blockarray}{cccccc}
\ & \text{\scriptsize $[p_1]$} & \text{\scriptsize $[p_2]$} & \text{\scriptsize [...]}  & \text{\scriptsize $[p_m]$}\\
\begin{block}{c(ccccc)}
\text{\scriptsize{$[D_1]$}} & h_{11} & h_{12} & ... & h_{1m} \\
\text{\scriptsize{$[D_2]$}} & h_{21} & h_{22} & ... & h_{2m} \\
\text{\scriptsize{$[...]$}} & ... & ... & ... & ... \\
\text{\scriptsize{$[D_k]$}} & h_{k1} & h_{k2} & ... & h_{km} \\
\end{block}
\end{blockarray} \ \rm,
\\
&m_c=\begin{blockarray}{ccccccc}
\ & \text{\scriptsize $[p_1]$} & \text{\scriptsize $[p_2]$} & \text{\scriptsize [...]}  & \text{\scriptsize $[p_{m-1}]$} & \text{\scriptsize $[p_m]$}\\
\begin{block}{c(cccccc)}
\text{\scriptsize{$[p_1]$}} & 0 & c_{12} & ...& c_{1(m-1)} & c_{1m} \\
\text{\scriptsize{$[p_2]$}} & c_{12} & 0 & ...& c_{2(m-1)} & c_{2m} \\
\text{\scriptsize{$[...]$}} & ... & ... & ... & ...& ... \\
\text{\scriptsize{$[p_{(m-1)}]$}} & c_{1(m-1)} & c_{2(m-1)} & ... & 0 & c_{(m-1)m} \\
\text{\scriptsize{$[p_{m}]$}} & c_{1m} & c_{2m} & ... & c_{(m-1)m} & 0 \\
\end{block}
\end{blockarray} \ \rm,
\end{split}
$}
\end{equation}

where $m_b$ is the bit-check adjacency matrix, $m_p$ is the phase-check adjacency matrix and $m_c$ is the cross-check adjacency matrix. In the above, the rows and columns are labelled to indicate whether they refer to data qubits $D$ (triangles in the operational representation) or parity qubits $P$ (stars in the operational representation). The binary variables, $b$, $h$ and $c$, indicate error propagation between two qubits if set to `1'. Note that the cross-check matrix is always symmetric around the diagonal to account for the fact that phase errors propagate in both directions. As an example, the adjacency matrices for the $[[4,2,2]]$ code depicted in figure \ref{fig:422_app} are given by
\begin{equation} \label{eq:adj_mat}
\begin{split}
&m_b=\begin{blockarray}{cccccc}
\ & \text{\scriptsize $[p1]$} & \text{\scriptsize $[p2]$}  \\
\begin{block}{c(ccccc)}
\text{\scriptsize{$[D_1]$}} & 1 & 0  \\
\text{\scriptsize{$[D_2]$}} & 1 & 0   \\
\end{block}
\end{blockarray},
\ \
m_p=\begin{blockarray}{cccccc}
\ & \text{\scriptsize $[p1]$} & \text{\scriptsize $[p2]$}  \\
\begin{block}{c(ccccc)}
\text{\scriptsize{$[D_1]$}} & 0 & 1  \\
\text{\scriptsize{$[D_2]$}} & 0 & 1   \\
\end{block}
\end{blockarray},
\\
&m_c=\begin{blockarray}{cccccc}
\ & \text{\scriptsize $[p1]$} & \text{\scriptsize $[p2]$}  \\
\begin{block}{c(ccccc)}
\text{\scriptsize{$[p1]$}} & 0 & 1  \\
\text{\scriptsize{$[p2]$}} & 1 & 0   \\
\end{block}
\end{blockarray}\rm.
\end{split}
\end{equation}		
From the $m_b$ matrix above, we can see that bit-errors on data qubits $D_1$ and $D_2$ are detected by parity qubit $P_1$. This is consistent with the black edges in figure \ref{fig:422_app}. Likewise, from $m_p$, we see that phase errors are detected by parity qubit $p_2$. Finally, adjacency matrix $m_c$ tells us that phase-errors on qubit $p_1$ are detected by qubit $p_2$ and vice-versa.

The GF(2) quantum parity matrix $G_{\footnotesize XZ}({\mathcal{S}_{\rm CPC}})$ for a CPC code is written in terms of its adjacency matrices as follows
\begin{equation} \label{eq:fin_geo_stabs}
\begin{split}
	&G_{\footnotesize XZ}({\mathcal{S}_{\rm CPC}})=\\
	&\begin{blockarray}{ccccccccccccccccccc}
\ & \text{\scriptsize $[D_1,...,D_k]$} & \text{\scriptsize $[p_1,...,p_{n-k}]$} & \text{\scriptsize $[D_1,...,D_k]$} & \text{\scriptsize $[p_1,...,p_{n-k}]$} \\
\begin{block}{c(cc|cccccccccccccccc)}
& m_p^T & m_b^T\cdot m_p\oplus m_c & m_b^T  & \openone_{n-k}  \\
\end{block}
\end{blockarray}\rm.
\end{split}
\end{equation}
The stabilizers $\mathcal{S}_{\rm CPC}$ of the CPC code are given by the rows of the quantum parity matrix. As an example, consider the quantum parity check matrix for the $[[4,2,2]]$ CPC code obtained by substituting equation (\ref{eq:adj_mat}) into equation (\ref{eq:fin_geo_stabs})
\begin{equation} \label{eq:fin_geo_422_stabs}
\begin{split}
&G_{\footnotesize XZ}({\mathcal{S}_{\rm [[4,2,2]]}})=\\&
\begin{blockarray}{ccccccccccccccccccc}
\ & \text{\scriptsize $[D_1]$} & \text{\scriptsize $[D_2]$} & \text{\scriptsize $[p_1]$} & \text{\scriptsize $[p_2]$} & \text{\scriptsize $[D_1]$} & \text{\scriptsize $[D_2]$} & \text{\scriptsize $[p_1]$} & \text{\scriptsize $[p_2]$} \\
\begin{block}{c(cccc|cccccccccccccc)}
& 0 & 0 & 0 & 1 & 1 & 1 & 1 & 0\\
& 1 & 1 & 1 & 0 & 0 &0 & 0 & 1 \\
\end{block}
\end{blockarray}\rm.
\end{split}
\end{equation}
From the above, the two GF(2) rows translate to the stabilizers $Z_{D_1}Z_{D_2}Z_{p_1}X_{p_2}$ and $X_{D_1}X_{D_2}X_{p_1}Z_{p_2}$ in Pauli notation.

The stabilizers of a quantum code must mutually commute. This condition means that the quantum parity check matrix of the code must satisfy the following relation
\begin{equation} \label{eq:symplectic}
G_X \cdot G_Z^T \oplus G_Z \cdot G_X^T =0 \rm,
\end{equation}
where $G_X$ and $G_Z$ are the $X$- and $Z$- components of the quantum parity check matrix given by $G_{XZ}=(G_X \ | \ G_Z)$. Substituting equation (\ref{eq:fin_geo_stabs}) into equation (\ref{eq:symplectic}) we obtain
\begin{equation}\resizebox{\columnwidth}{!}{$\begin{split}
&G_X G_Z^T\oplus G_Z G_X^T =(m_p^T  \ | \ m_b^T\cdot m_p\oplus m_c ) \cdot (m_b^T  \ | \  \openone_{n-k})^T \\& \oplus (m_b^T  \ | \  \openone_{n-k}) \cdot (m_p^T  \ | \ m_b^T\cdot m_p\oplus m_c )^T\\
&=(m_p^T \cdot m_b \ | \  m_b^T\cdot m_p\oplus m_c) \oplus (m_b^T \cdot m_p \ | \ m_p\cdot m_p^T \oplus m_c^T )\\&= (m_p^T \cdot m_b \oplus m_p^T \cdot m_b \ | \   m_b^T\cdot m_p \oplus m_c \oplus m_b^T\cdot m_p \oplus m_c^T) \\&=0
\end{split}$}
\end{equation}
From the above, we see that the CPC code structure guarantees a set of commuting stabilizers for all combinations of the adjacency matrices $m_b$, $m_p$ and $m_c$. As such, the CPC framework provides a method for creating a valid stabilizer code from any sequence of parity checks.

\bibliographystyle{IEEEtran}

\bibliography{factor-bib}

\begin{IEEEbiographynophoto}{Joschka Roffe}
is a research associate at The University of Sheffield. He graduated with a physics (MPhys) degree from The University of Manchester in 2015. Following this, he studied a PhD in quantum computing at Durham University under the supervision of Viv Kendon. His thesis, `The Coherent Parity Check Framework for Quantum Error Correction', was awarded the 2019 Durham Winton Doctoral Prize in Computational Physics. Joschka now works as part of the Quantum Code Design and Architectures project (\url{http://www.qcda.eu}). His current research focuses on the design and implementation of quantum error correction protocols. Up-to-date information about Joschka can be found on his website: \url{http://www.roffe.eu}.
\end{IEEEbiographynophoto}

\begin{IEEEbiographynophoto}{Stefan Zohren}
is an Associate Professorial Research Fellow at Machine Learning Research Group and the Oxford-Man Institute for Quantitative Finance, University of Oxford. Previously, he coordinated the Quantum Optimisation and Machine Learning project, a joined research project of Oxford University, Nokia Technologies and Lockheed Martin. His background is in theoretical physics, probability theory and statistics. Stefan’s research interests include machine learning applied to finance, deep learning for time series modelling as well as quantum computing and statistical physics.
\end{IEEEbiographynophoto}

\begin{IEEEbiographynophoto}{Dominic Horsman}
Dominic Horsman is Chair of Excellence in Quantum Engineering at the University of Grenoble, and previously worked on the NQIT quantum technologies project in the UK. He has a background in theoretical physics and computer science, as well as industry experience in AI. His research focus is developing the foundations of quantum software (including novel language tools based on the ZX-calculus), error correction, and compilation. His work is directed towards the development of both near-term (NISQ) quantum hardware and longer-term full scale quantum computers.
\end{IEEEbiographynophoto}

\begin{IEEEbiographynophoto}{Nicholas Chancellor}
is an EPSRC UKRI Innovation fellow at Durham University (UK). He did his PhD. at the University of Southern California in 2013 under the supervision of professor Stephan Haas. Prior to becoming a principle investigator on his own grant in 2018, he was postdoc at University College London and Durham University. Nicholas has 20 publications either accepted or published in peer reviewed journals. His main subject of research is continuous time quantum computing including quantum annealing, in particular hybrid quantum-classical algorithms, where he helped pioneer the use of reverse annealing as an algorithmic tool. Nicholas has also helped to develop the coherent parity check (CPC) formalism for quantum error correction. Up-to-date information about Nicholas can be found on his personal webpage \url{http://www.nicholas-chancellor.me}.
\end{IEEEbiographynophoto}

\end{document}